\documentclass[preprint,amsmath,amssymb, aps, prl,floatfix,nofootinbib]{revtex4-1} 
\usepackage{graphicx}
\usepackage{dcolumn}
\usepackage{bm}
\usepackage{caption}
\usepackage{amssymb,amsmath,color}
\usepackage{url}
\usepackage{float}
\usepackage{braket}
\usepackage{hyperref} 
\usepackage{array}
\usepackage{comment}% Works
\hypersetup{colorlinks = true, linkcolor = blue, anchorcolor = blue, citecolor = blue, filecolor = blue, urlcolor = blue}
\usepackage{xcolor}

\usepackage[T1]{fontenc}
\usepackage[utf8]{inputenc}

\begin{document}
\title{Atomistic theory of twist-angle dependent intralayer and interlayer exciton properties in twisted bilayer materials}

\author{Indrajit Maity} 
\author{Arash A. Mostofi}
\author{ Johannes Lischner} 
\email{j.lischner@imperial.ac.uk}
\affiliation{Departments of Materials and Physics and the Thomas
Young Centre for Theory and Simulation of Materials, Imperial
College London, South Kensington Campus, London SW7 2AZ, UK}

\date{\today}
\maketitle

Twisted bilayers of two-dimensional materials have emerged as a highly tunable platform to study and engineer properties of excitons. However, the atomistic description of these properties has remained a significant challenge as a consequence of the large unit cells of the emergent moir\'e superlattices. To address this problem, we introduce an efficient atomistic quantum-mechanical approach to solve the Bethe-Salpeter equation that exploits the localization of atomic Wannier functions. We then use this approach to study intra- and interlayer excitons in twisted  WS$_{2}$/WSe$_{2}$ at a range of twist angles. In agreement with experiment, we find that the optical spectrum exhibits three low-energy peaks for twist angles smaller than $2^\circ$. The energy splitting between the peaks is described accurately. We also find two low-energy interlayer excitons with weak oscillator strengths. Our approach opens up new opportunities for the atomistic design of light-matter interactions in ultrathin materials.  

\section{Introduction}
The stacking and twisting two (or more) layers of two-dimensional (2D) materials results in the emergence of a large-scale moir\'{e} pattern which gives rise to novel properties. Recently, such moir\'{e} materials have attracted tremendous interest as a platform to explore exotic optical properties~\cite{Reganemerging2022, Jianginterlayer2021}. For example, intralayer excitons (where the constituent electron and hole reside in the same layer) exhibit topological properties~\cite{Fengchengtopological2017} and a splitting of the lowest-lying bright exciton at small twist angles~\cite{Fengchengtopological2017, Jinobservation2019, Andersenexcitons2021, Sandhyahyper2022}. On the other hand, interlayer excitons (where the constituent electron and hole reside in different layers) show remarkably long lifetimes up to a few hundred of nanoseconds~\cite{Riveravalley2016, Jonghwanobservation2017, Choitwist2021, Yuantwist2020, Rossiphason2023}, which makes them promising candidates for realizing exotic quantum phenomena, such as superfluidity~\cite{Foglerhigh2014, Niclasmoire2022, Camillekey2021}, supersolidity~\cite{Aleksinonlocal2022} and Bose-Einstein condensation~\cite{Wangevidence2019}, and for designing efficient excitonic devices~\cite{Unuchekroom2018, Riveravalley2016}.

To gain a microscopic understanding of the properties of excitons in moir\'e materials and how these properties depend on the twist angle, strain, electric and magnetic fields and the composition of the individual layers, theoretical calculations can play an important role. To date, most theoretical studies of excitons in moir\'e materials are based on the effective mass approximation~\cite{Fengchengtopological2017, Bremtunable2020}. These studies have provided many important insights, but it is challenging to incorporate important factors, such as atomic relaxations or multi-valley effects. In principle, such effects can be straightforwardly captured using first-principles techniques, such as the ab initio density functional theory (DFT) combined with many-body perturbation theory (GW) and the Bethe-Salpeter equation (BSE) approach. However, application of this approach to twisted bilayer systems is numerically extremely challenging because of its unfavourable scaling with system size of this approach~\cite{Kunduexciton2023}. As a consequence, additional simplifications are usually introduced~\cite{Honglishedding2020}. For example, computed monolayer spectra are interpolated to approximate the optical properties of the twisted bilayer~\cite{Tranevidence2019, Elyseoptical2022} or the single-particle wavefunctions of the moir\'{e} system are approximated in terms of monolayer wavefunctions ~\cite{Naikintralayer2022}. However, a full solution of the BSE for a twisted bilayer system has not yet been achieved at small twist angles. The intriguing discovery of intralayer charge transfer excitons by Naik and coworkers~\cite{Naikintralayer2022} emphasizes the importance of accurate atomistic models for capturing excitons in moiré materials, as these excitons cannot be captured using continuum models based on effective mass approximations.

In this paper, we introduce an atomistic approach to solve the BSE which exploits the localization of Wannier functions. Using this approach, we investigate the dependence of intralayer exciton properties on the twist angle in WS$_{2}$/WSe$_{2}$, a prototypical transition metal dichalcogenide heterobilayer. In agreement with experiments~\cite{Jinobservation2019, Rossinterlayer2017,Yueall2023}, we find that the low-energy intralayer exciton peak splits into three peaks at small twist angles, and we observe good agreement with the available experimental data. We also analyze the wavefunction of the low-energy intralayer excitons and find that the character of the second lowest exciton depends sensitively on twist angle. We also observe interlayer excitons with an excitation energy of approximately 1.4 eV. 

%IM: 14/06/2024
\textit{Bethe-Salpeter Equation in Wannier basis.} To study properties of excitons in twisted bilayer WS$_2$/WSe$_2$, we solve the Bethe-Salpeter equation (BSE) for the interacting electron-hole Green's function~\cite{Salpeterrelativistic1951, Rohlfingelectron2000}. For zero-momentum excitons of relevance to optical absorption, the
BSE Hamiltonian is given by
\begin{equation}
\begin{split}
& \langle cv{\bf k}| \hat{H}_{\text{BSE}} |c^{\prime}v^{\prime}{\bf k^{\prime}}\rangle =  \\
& (\epsilon_{c{\bf k}} - \epsilon_{v{\bf k}}) \delta_{cc^\prime} \delta_{vv^\prime} \delta_{\bf kk^\prime} - D_{cv{\bf k}, c^{\prime}v^{\prime}{\bf k^{\prime}}} + X_{cv{\bf k}, c^{\prime}v^{\prime}{\bf k^{\prime}}},
\label{bse}
\end{split}
\end{equation}
where $\epsilon_{c(v){\bf k}}$ denotes the quasiparticle energy of an electron in conduction (valence) band $c$ ($v$) with crystal momentum ${\bf k}$ with wavefunction $\psi_{c(v){\bf k}}$. Also, $D$ and $X$ are the direct and exchange interaction terms~\cite{Rohlfingelectron2000, Fengchengexciton2015} given by
\begin{equation}
\begin{split}
& D_{cv{\bf k}, c^{\prime}v^{\prime}{\bf k^{\prime}}} = \\
& \int d{\bf x} d{\bf x^\prime} \psi^{*}_{c,\bf{k}}({\bf x}) \psi_{c^\prime,\bf{k^\prime}}({\bf x}) W({\bf r, r^\prime}) \psi_{v{\bf k}} ({\bf x^\prime}) \psi^{*}_{v^\prime{\bf k^\prime}} ({\bf x^\prime}),\\
& X_{cv{\bf k}, c^{\prime}v^{\prime}{\bf k^{\prime}}} \\
& = \int d{\bf x} d{\bf x^\prime} \psi^{*}_{c,\bf{k}}({\bf x}) \psi_{v,\bf{k}}({\bf x}) V({\bf r, r^\prime}) \psi_{c^\prime{\bf k^\prime }} ({\bf x^\prime}) \psi^{*}_{v^\prime{\bf k}^{\prime}} ({\bf x^\prime}),
\end{split}
\end{equation}
where ${\bf x}=({\bf r},s)$ is a composite index representing both spatial and spin degrees of freedom, and $W({\bf r, r^\prime})$ and $V({\bf r, r^\prime})$ are the screened and bare Coulomb interactions, respectively. The evaluation of these integrals for large systems is computationally challenging, in particular for standard implementations that employ a plane-wave expansion of the single-particle wavefunctions and the Coulomb interactions~\cite{Deslippeberkeleygw2012}.

%IM: 14/06/2024
To overcome this problem, we expand the single-particle wavefunctions in a basis of Wannier functions~\cite{Marzarimaximally2012} and then exploit the localization of these basis functions to efficiently construct the BSE Hamiltonian. In particular, we first perform an ab initio DFT calculation for the twisted bilayer system and then construct Wannier functions using the Wannier90 code~\cite{Pizziwannier902020}. We include five $d$-like Wannier functions for metal atoms and three $p$-like Wannier functions for chalcogen atoms (note that spin-orbit coupling is included after the Wannier function generation, see SI~\cite{SI}, Sec.~C for details). The DFT Bloch states $\psi_{m{\bf k}}$ can be expanded in terms of the Wannier function $\phi_n$ as 
\begin{equation}
\psi_{m{\bf k}}({\bf r}) = \frac{1}{\sqrt{N}} \sum_{\bf R} e^{i{\bf k}\cdot{\bf R}} \sum_{n} C^{{\bf k}}_{nm} \phi_{n} ({\bf r}-{\bf R} - {\bf t}_{n}),
\end{equation}
where ${\bf R}$ denotes a lattice vector, $C^{\bf k}_{nm}$ are the expansion coefficients, ${\bf t}_{n}$ represents the position of $\phi_{n}$ in the home unit cell at ${\bf R}=0$ and $N$ is the number of ${\bf k}$ points used to sample the moir\'{e} Brillouin zone. More details of the Wannier function generation can be found in the Methods section. We provide evidence for the completeness of our Wannier function basis by comparing the band structures obtained from DFT with the one obtained from diagonalizing the DFT Hamiltonian in the Wannier function basis, see Sec.~C of the SI~\cite{SI}. The Kohn-Sham band gaps of monolayer WS$_{2}$ and WSe$_{2}$ are corrected using the GW method through rigid shifts~\cite{Filipcomputational2015}. 

%IM: 14/06/2024
In the Wannier function basis, the BSE Hamiltonian can be approximated by using the localized and orthogonal nature of Wannier functions, along with the translational invariance of Coulomb interactions
\begin{widetext}
\begin{equation}
\begin{split}
& \langle cv{\bf k}| \hat{H}_{\text{BSE}} |c^{\prime}v^{\prime}{\bf k^{\prime}}\rangle \\
& = (\epsilon_{c{\bf k}} - \epsilon_{v{\bf k}}) \delta_{cc^\prime} \delta_{vv^\prime} \delta_{\bf kk^\prime} -\frac{1}{N} \sum_{{\bf R},n_{1},n_{3}} {(C^{\bf k}_{n_{1}c}})^* C^{\bf k^\prime}_{n_{1}c^\prime} {C^{\bf k}_{n_{3}v}} {(C^{\bf k^\prime}_{n_{3}v^\prime})}^* W({\bf R} + {\bf t}_{n_3} - {\bf t}_{n_1}) e^{i ({\bf k - k^\prime})\cdot{\bf R}} \\
& \hspace{1.5 in} + \frac{1}{N} \sum_{{\bf R},n_{1},n_{3}} {(C^{\bf k}_{n_{1}c})}^* {C^{\bf k}_{n_{1}v}} C^{\bf k^\prime}_{n_{3}c^\prime} {(C^{\bf k^\prime}_{n_{3}v^\prime})}^* V({\bf R} + {\bf t}_{n_{3}} - {\bf t}_{n_{1}}).
\label{bsewannier}
\end{split}
\end{equation}
\end{widetext}
A detailed derivation of this expression is provided in Sec.~B of the SI along with a discussion of all approximations that have been made.

%IM: 14/06/2024
To determine the screened interaction $W$ in a twisted bilayer system, we consider each layer as a polarizable sheet and solve the resulting electrostatic problem~\cite{Cudazzodielectric2011, Keldyshcoulomb1979, Berkelbachtheory2013, Ridolfiexcitonic2018,Danovichlocalized2018, Aghajanianoptical2023}, see Sec.~C of the SI~\cite{SI}. When the electron and the hole reside on the same layer, the screened interaction is given by 
\begin{equation}
W(r) = -\frac{1}{4\pi\epsilon_{0}\epsilon_{\text{bg}}} \frac{e^{2}\pi}{2r_{s}} [ H_{0}(r/r_{s})-Y_{0}(r/r_{s})], 
\end{equation}
where $r_s=2\pi (\alpha_1 + \alpha_2)$ is a characteristic length scale determined by the polarizabilities $\alpha_i$ of the two layers~\cite{Berkelbachtheory2013}, $\epsilon_{\text{bg}}$ is the background dielectric constant due to the presence of the substrate, and $H_0$ and $Y_0$ are the zeroth-order Struve and Bessel functions of the second kind, respectively. 

%IM: 14/06/2024
When the electron and hole reside on different layers, the screened interaction is given by the same expression as above, but $r_s$ is replaced by $r_s + d$ with $d$ being the average interlayer distance. We use $\alpha_{\mathrm{WS_2}}=6.03$ \AA\, $\alpha_{\mathrm{WSe_2}}=7.18$ \AA\, and $d=7$ \AA\ to represent the polarizabilities of the WS$_{2}$ and WSe$_{2}$ (calculated from the inverse dielectric tensor using DFT in conjunction with random-phase-approximation~\cite{Berkelbachtheory2013}) and the average interlayer separation of the heterobilayer, respectively. 

%IM: 14/06/2024
The optical conductivity of the twisted bilayer WS$_{2}$/WSe$_{2}$ is given by
\begin{equation}
\operatorname{Re}[\sigma_{xx}(\omega)] \propto \sum_{S} \left|\sum_{cv{\bf k}} A^{S}_{cv{\bf k}} \langle v{\bf k}|p_{x}|c{\bf k}\rangle\right|^{2} \delta(\omega - \omega^{S}),
\label{OptCond}
\end{equation}
where $\omega^{S}$, and $A^{S}_{cv{\bf k}}$ denote the eigenvalues and eigenvectors of the BSE Hamiltonian, respectively, and the matrix elements of the momentum operator are expressed as $\langle v{\bf k}|{\bf p}|c{\bf k}\rangle=\sum_{n_{1}n_{2}} (C^{\bf k}_{n_{2}v})^{*} C^{\bf k}_{n_{1}c} (\nabla_{\bf k} H^{\bf k}_{n_{1}n_{2}} +  i ({\bf t}_{n_{1}} - {\bf t}_{n_{2}}) H^{\bf k}_{n_{1}n_{2}})$ with $H^{\bf k}$ being the Hamiltonian at $\mathbf{k}$ in the Wannier function basis. The effect of spin-orbit coupling is included as a perturbation~\cite{Dianascreening2016}. More details can be found in the SI, Sec.~C~\cite{SI}. 

%IM: 14/06/2024
The exciton wavefunction for a fixed hole position $\mathbf{r}_h^{0}$ is obtained from
\begin{equation}
\begin{split}
& |\Psi^{S}({\bf r}_{e}={\bf R}_{j_{1}}+{\bf t}_{n_{1}}, {\bf r}_{h}={\bf r}_{h}^{0})|^{2}\newline \propto |\sum_{vc{\bf k}} A^{S}_{vc {\bf k}} \times \\
&  \sum_{n_{3}\in{\bf r}_{h}} (C^{\bf k}_{n_{3}v})^*\times e^{i{\bf k}\cdot{\bf R}_{j_{1}}} C^{\bf k}_{n_{1}c}|^{2}
\end{split}
\end{equation}
with $n_{3}\in{\bf r}_{h}^{0}$ denoting the set of orbitals centered at $\mathbf{r}_h^{0}$.

%IM: 14/06/2024
To assess the accuracy of our approach, we apply it to monolayer WS$_{2}$ and WSe$_{2}$. We find good agreement with experimental results and plane-wave BSE calculations at a significantly reduced computational cost, see Sec.~D of the SI~\cite{SI}. We also note that similar approaches have recently been used to compute optical properties of monolayers of MoS$_{2}$~\cite{EfficientAlejandro2024,Alexandrewantibexos2023}. Below, we apply our approach to study low-energy intralayer and interlayer excitons in a WS$_{2}$/WSe$_{2}$ heterobilayer.

\section{Results}

\subsection{Intralayer excitons} 
\begin{figure*}[ht]
    \centering
    \includegraphics[scale=0.7]{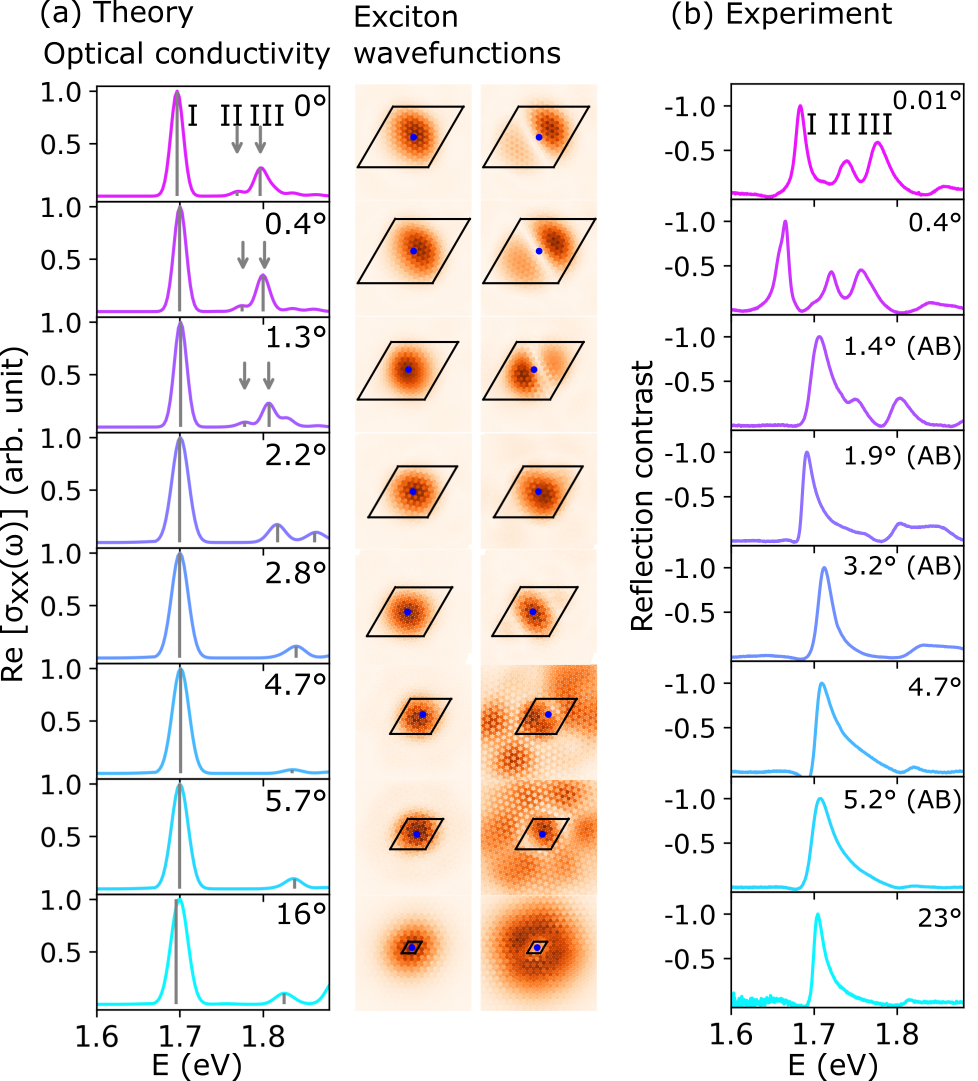}
    \caption{(a): Optical conductivity (left panel) and wavefunctions of the two lowest-energy intralayer excitons (right panel) of twisted bilayer WS$_{2}$/WSe$_{2}$ at a range of twist angles. Grey arrows indicate the energies of the excitons and blue dots indicate fixed hole positions at the center of the home unit cell. The optical conductivities are normalized such that the height of the main peak is unity. Additionally, the main peak is shifted to match the main peak of the monolayer spectrum. (b): Experimental reflection contrast spectrum from Ref.~\cite{Jinobservation2019}.  When experimental data for specific twist angles near 0$^\circ$ is not available, we compare to the corresponding twist angles near 60$^\circ$ (indicated by AB) since intralayer excitons at 60$^\circ + \theta$ have similar properties as those at $\theta$~\cite{Jinobservation2019}.}
   \label{fig1}
\end{figure*}

\begin{figure}[ht]
    \centering
    \includegraphics[scale=0.7]{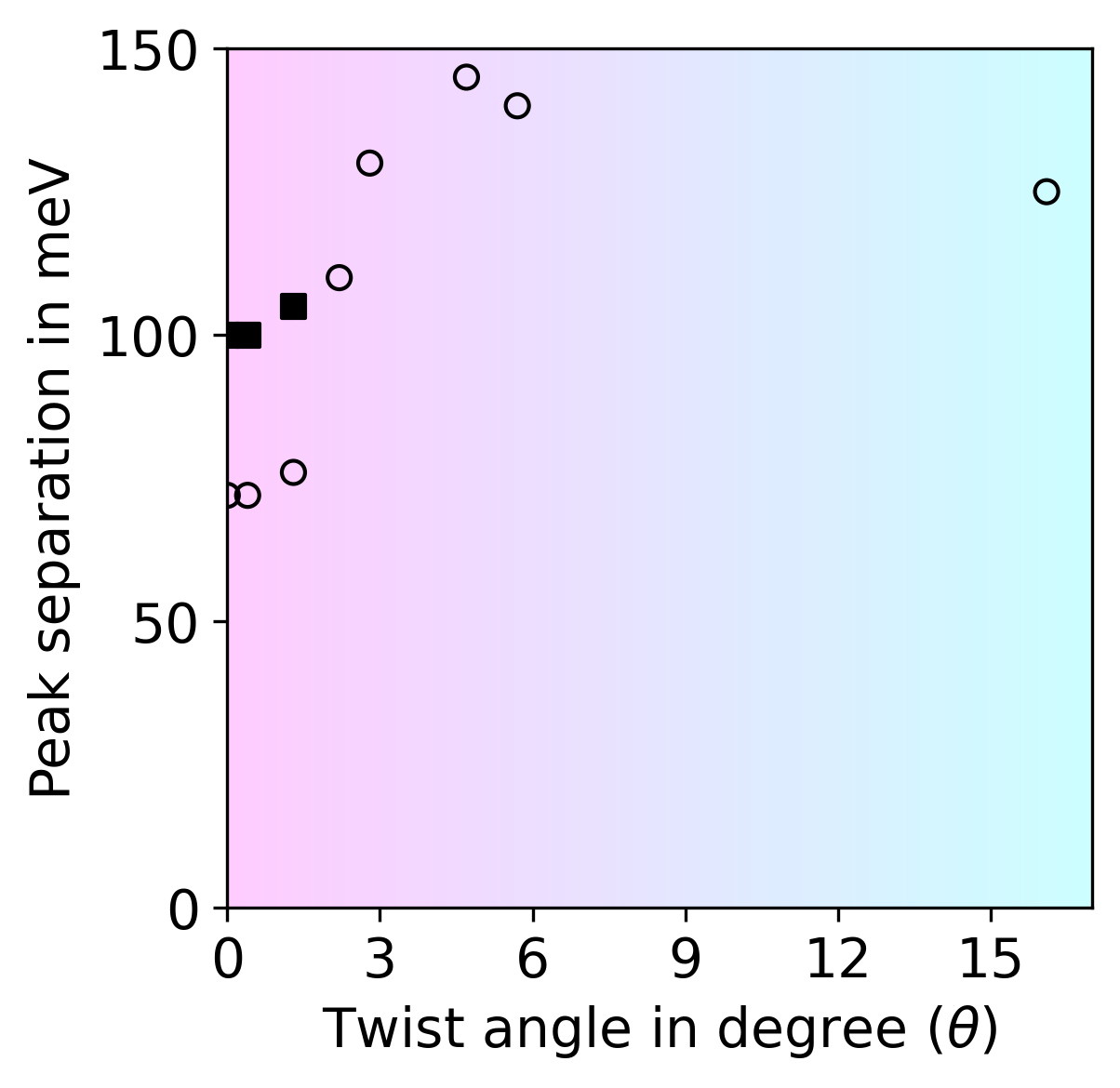}
    \caption{Energy separation between peaks I and II (black dots) and peaks I and III (black squares) in the optical conductivity (shown in Fig.~\ref{fig1}) of twisted bilayer WSe$_2$/WS$_2$ as function of twist angle.}
   \label{fig2}
\end{figure}

Figure~\ref{fig1} compares the computed optical conductivity from low-energy intralayer excitons in twisted WS$_2$/WSe$_2$ to the measured reflection contrast spectrum~\cite{Jinobservation2019} for a range of twist angles. Low-energy intralayer excitons are localized in the WSe$_2$ layer and this allows us to consider only the relaxed WSe$_2$ layer in the BSE calculation, see Methods for details. As the twist angle is reduced, the computed spectra exhibit several significant changes: (i) at twist angles below $\sim 2^\circ$, the spectrum exhibits three peaks (labelled I, II, and III) in the low-energy region, (ii) the spectral weight contained in the additional peaks II and III increases at small twist angles relative to the spectral weight in peak I, and (iii) the energy separation between peaks I and II and between peaks I and III shrinks as the twist angle is reduced, as shown in Fig.~\ref{fig2}. These findings are in qualitative agreement with the experimental observations. In particular, the peak separations at small twist angles agree within a few meV. However, our approach underestimates the spectral weight contained in peak II. 

Figure~\ref{fig1}(a) also shows the exciton wavefunctions of the two lowest-energy intralayer excitons for a fixed hole position at the center of the home unit cell. For all twist angles, the lowest-energy intralayer exciton has a hydrogenic 1s character. Near a twist angle of zero degree, however, the distribution of the electron around the hole becomes elliptical. In contrast, the character of the exciton which produces the second peak undergoes several qualitative changes as the twist angle is reduced. For the largest twist angle (e.g., 16$^\circ$), the second peak is produced by a 2p hydrogenic state. At intermediate twist angles (5.7$^\circ$ and 4.7$^\circ$), the peak is instead generated by an exciton with mixed 2s and 2p character. At intermediate twist angles (e.g., 2.2 $^\circ$), the wavefunction associated with the second peak looks similar to the lowest-energy 1s state. We interpret this as a folded finite-momentum state. Finally, at very small twist angles (e.g., 0.4$^\circ$), the state producing the second peak again has a 2p-like character. The full evolution of the spectrum of all low-energy excitons is shown in the SI, Sec.~H.

This evolution of the character of the intralayer exciton as function of twist angle is a consequence of the interplay between electron-hole interactions and the moiré potential: at large twist angles, the moiré potential is negligibly small and the exciton properties are determined by electron-hole interactions which only depend on the relative distance between electron and hole. At small twist angles, the moiré potential is large and determines the region where the exciton localizes.

\begin{figure*}[ht]
    \centering
    \includegraphics[scale=0.7]{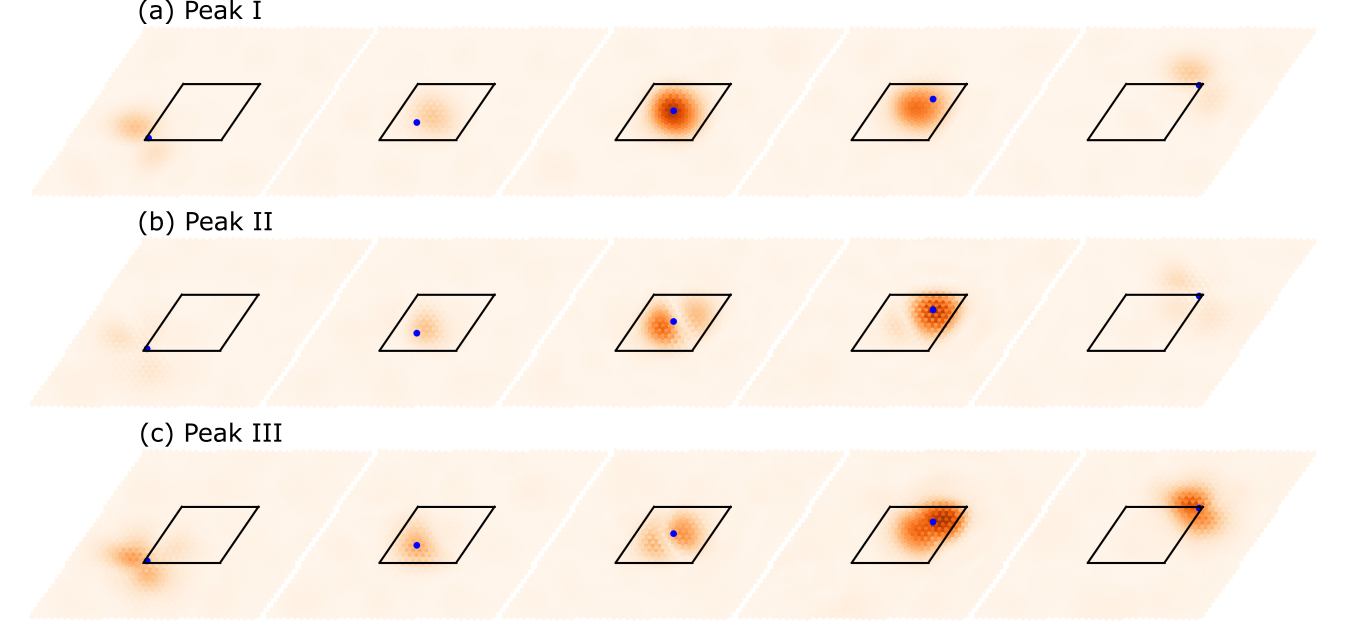}
    \caption{Wavefunctions of the low-energy intralayer excitons in 1.3$^\circ$ twisted WS$_{2}$/WSe$_{2}$ for different hole positions indicated by a blue dot. The home unit cell is marked with black solid lines.}
   \label{fig4}
\end{figure*}

In order to demonstrate the importance of the moir\'e potential, the electron distribution for different positions of the hole is examined at a small twist angle, see Fig.~\ref{fig4}. When the hole is fixed in certain regions of the moiré unit cell (e.g. at the corner of the unit cell), the exciton wavefunction is small in magnitude for all values of the electron coordinate. In general, we find that the electron ``follows'' the hole (i.e. the exciton has a Wannier-type character), but the shape of the distribution changes. In contrast to Naik and coworkers~\cite{Naikintralayer2022}, we do not observe the exciton corresponding to peak III to have significant intralayer charge-transfer character. This is likely a consequence of the smaller unit cell used in our calculations which gives rise to a weaker moir\'e potential, see SI, Sec. F~\cite{SI} for a comparison of the Kohn-Sham wavefunctions, where we show that the electron and hole wavefunctions always have overlap at relevant stackings.

\begin{figure}[ht!]
    \centering
    \includegraphics[scale=0.55]{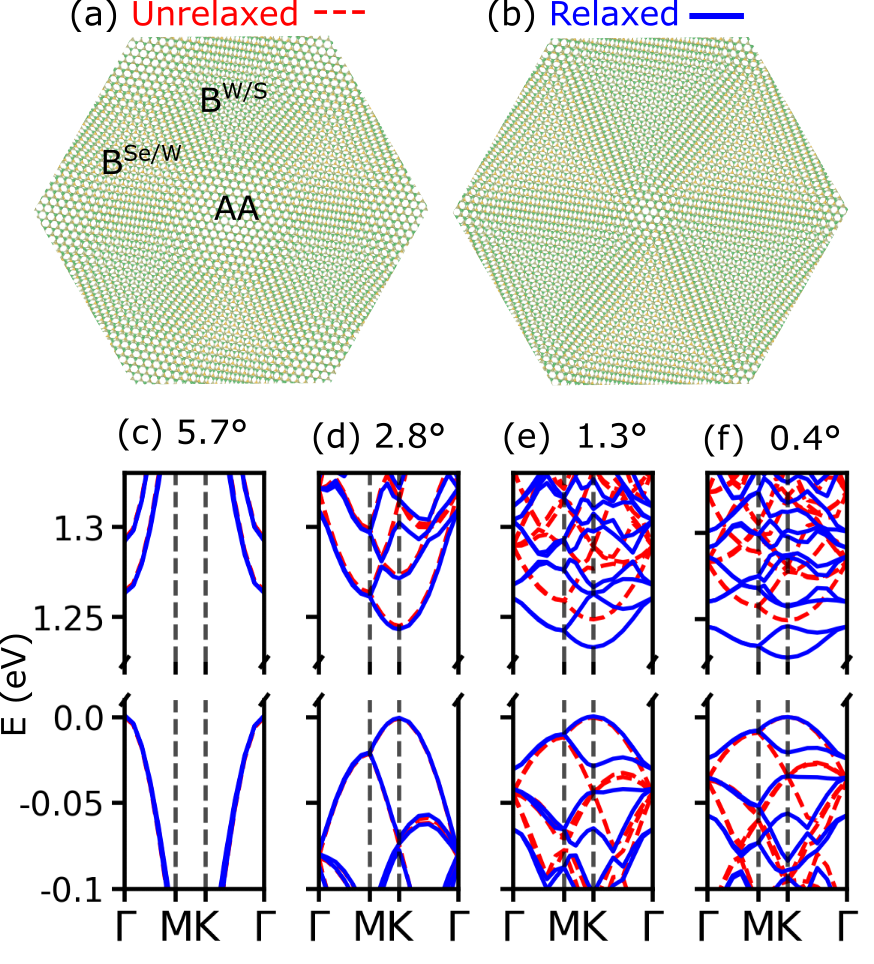}
    \caption{(a),(b): Atomic relaxations lead to the shrinking of regions with unfavorable AA stacking and the growth of regions with low-energy $\mathrm{B^{W/S}}$ and $\mathrm{B^{Se/W}}$ stacking in $\mathrm{WS_{2}/WSe_{2}}$. The high-symmetry stackings are labeled. (c)-(f): Electronic band structures of isolated $\mathrm{WSe_{2}}$ for several twist angles obtained from ab initio density functional theory calculations that include spin-orbit coupling. The WSe$_{2}$ layer was isolated after performing structural relaxation of the WS$_{2}$/WSe$_{2}$ heterobilayer. The blue solid lines indicate the results of the relaxed structures, while the red dashed lines indicate the results of the flat monolayer WSe$_{2}$ with the same dimensions.}
   \label{fig3}
\end{figure}

To understand the origin of the twist-angle dependent exciton properties in more detail, we analyze the atomic structure of the twisted bilayer and its electronic band structure, see Fig.~\ref{fig3}. We find that atomic relaxations play an important role at twist angles smaller than $2^\circ$. For example, we find that the size of the AA stacking regions shrinks by 4\% at a twist angle of 5.7$^\circ$ after atomic relaxations. However, at a twist angle of 1.3$^\circ$, the size of these regions shrinks by 50\%, see Figs.~\ref{fig3}(a) and (b). See Methods for additional details on the estimation of the AA stacking area.

Atomic relaxations have an important effect on the electronic band structure, see Figs.~\ref{fig3}(c)-(f). Here we compare the band structures of the WSe$_{2}$ layer (after relaxation and removal of the WS$_2$ layer) with those of a flat unrelaxed monolayer for several twist angles. At twist angles less than $2^\circ$, the band structure of the relaxed system deviates significantly from that of the unrelaxed system. Importantly, relaxations lead to a flattening of the highest valence band and also of the lowest conduction band and a reduction of the band gap as the twist angle decreases~\cite{Conradvisualizing2021}. In particular, the widths of the highest valence band and the lowest conduction band are reduced by $\sim 30\%$ at a twist angle of $1.3^\circ$ compared to the results for a flat unrelaxed monolayer.

\subsection{Interlayer excitons} 
\begin{figure}[ht]
    \centering
    \includegraphics[scale=0.55]{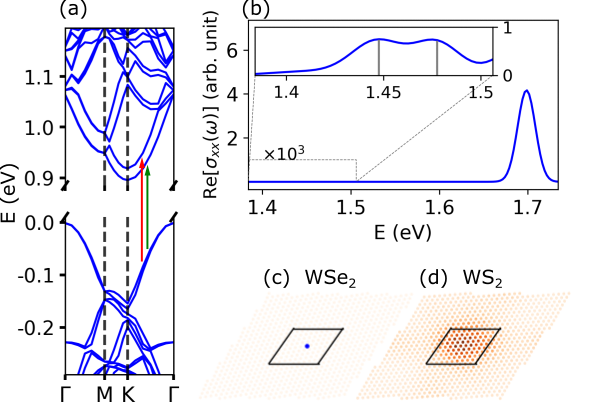}
    \caption{(a): Electronic band structure, (b): optical conductivity of the WS$_{2}$/WSe$_{2}$ twisted heterobilayer with a twist angle of 5.7$^\circ$. (c),(d): exciton envelope associated with the interlayer exciton (shown in the inset) with a hole fixed at the center of the home unit cell in the WSe$_{2}$ layer. The conduction spin-split bands in (a) give rise to two peaks in (b).}
   \label{fig5}
\end{figure}

We also study interlayer excitons of twisted WS$_{2}$/WSe$_{2}$. For this, we solve the BSE calculations for the fully relaxed bilayer. Fig.~\ref{fig5}(b) shows the optical conductivity of the twisted bilayer at a twist angle of $5.7^\circ$. Near 1.45~eV, two peaks with very small intensities are observed, see inset. These peaks arise from interlayer excitons in which the electron is localized on the WS$_2$ layer and the hole is localized on the WSe$_2$ layer, see Fig.~\ref{fig5}(c) and (d). The spatial separation of the electron and the hole results in a very small transition dipole moment and oscillator strength for these excitons~\cite{Fengchengtheory2018, Hongyianomalous2015, Rossinterlayer2017}. The calculated interlayer exciton energy is in agreement with previous experiments, with measured energies ranging from 1.35 to 1.45 eV~\cite{Jinobservation2019, Montblanchconfinement2021, Rossinterlayer2017, Yuantwist2020}.

The two peaks originate from transitions between spin-split partners of the CBM of WS$_{2}$ and VBM of WSe$_{2}$, see   Fig.~\ref{fig5}(a). The magnitude of the spin splitting is approximately 30 meV. From the electronic band structure of the twisted bilayer, it can be observed that the bandgap is indirect with the valence band maximum located at $\Gamma$ and the conduction band minimum at the $K$ point of the moir\'{e} Brillouin zone. This suggests that interlayer excitons with finite momenta can have lower energies than zero-momentum excitons. We leave the exploration of the Bethe-Salpeter equation for finite-momentum excitons to future work. 

\section{Discussion}
we have developed an efficient approach that exploits the localization of Wannier functions to solve the Bethe-Salpeter equation of twisted bilayer materials. We have applied this approach to study the properties of intra- and interlayer excitons in twisted WSe$_2$/WS$_2$. In agreement with experimental measurements, we find three peaks in the optical conductivity of intralayer excitons at small twist angles. Moreover, the energy separation between the peaks is well reproduced. The excitation energy of interlayer excitons is about 250 meV smaller than that of intralayer excitons, but their oscillator strengths are very weak. In the future, our approach can be straightforwardly applied to other multilayers of two-dimensional materials but also to other systems whose optical properties are difficult to describe with standard ab initio techniques, such as defects in solids or disordered materials. 

\section{Methods} \label{methods}
\subsection{A. Structure generation}
All the twisted WS$_{2}$/WSe$_{2}$ heterobilayer structures were generated using the TWISTER package~\cite{Naiktwister2022} using lattice constants of 3.32 \AA\ for monolayer WSe$_{2}$ and 3.18 \AA\ for WS$_{2}$, see  Supplementary Information (SI)~\cite{SI}, Sec.~A, for additional details. 

\subsection{B. Atomic relaxations}
We performed atomic relaxations using accurate classical interatomic potentials fitted to density functional theory calculations. We used the Stillinger-Weber potential to represent interactions within individual layers~\cite{Zhouhandbook2017} and a Kolmogorov-Crespi potential for the interactions between the layers ~\cite{Naikkolmogorov2019}. All relaxations are performed using the LAMMPS package~\cite{Thompsonlammps2022,lammps}. We used the FIRE algorithm~\cite{Bitzekstructural2006} to relax the atoms within a fixed simulation box with a force tolerance of $10^{-6}$ eV/\AA\ for any atom along any direction. 
To determine the size of the AA region in the relaxed and unrelaxed structures, we define the order parameter $\bf{u}$ as the minimum in-plane translation needed to transform any stacking configuration within the moiré pattern to AA stacking~\cite{Maityreconstruction2021}. We used $|{\bf u}|<0.5$ \AA\ as criterion to define the AA region. 

\subsection{C. Electronic structure calculations}
The electronic structure calculations were performed on both unrelaxed and relaxed structures using the SIESTA package that uses localized atomic orbitals as the basis~\cite{Solersiesta2002}. We used norm-conserving Troullier-Martins pseudopotentials~\cite{Troullierefficient1991} and the local density approximation to describe exchange-correlation effects~\cite{Perdewself1981}. We used a single-$\zeta$ plus polarization basis for the expansion of wavefunctions. For monolayer WSe$_{2}$, we also carried out calculations using a double-$\zeta$ plus polarization basis, but did not observe any significant differences in the resulting exciton properties, see SI~\cite{SI}, Sec.~E for details. For all calculations we used  $\Gamma$ point sampling to obtain the charge density and a plane-wave energy cutoff of 100 Rydberg. We used an energy shift of 0.02 Ry. A large vacuum spacing of $20$ \AA\ was used in the out-of-plane direction.

\subsection{D. Wannier function generation}
We used the relevant $d$ and $p$ orbitals as starting guess for the generation of Wannier functions via the one-shot projection method~\cite{Marzarimaximally2012}
as implemented in the Wannier90 code~\cite{Pizziwannier902020}. For example, we project the Kohn-Sham wavefunctions onto atom-centered $d$ ($d_{xy}, \ d_{yz}, \ d_{zx}, \ d_{x^2-y^2},\ d_{z^2}$) orbitals for every W atom, and $p$ ($p_{x}$, $p_{y}$, and $p_{z}$) orbitals for every S and Se atom and then orthogonalize. We have compared the resulting exciton properties to those obtained using maximally localized Wannier functions and found no significant differences, see SI~\cite{SI}, Sec.~E. A disentanglement procedure was used~\cite{Souzamaximally2001, Marzarimaximally2012}. We also compared our exciton properties to those obtained from Wannier functions generated from the plane-wave DFT code Quantum ESPRESSO package~\cite{Giannozziquantum2009, Giannozziadvanced2017}. Again, we find good agreement between the two approaches for monolayer WS$_{2}$ and WSe$_{2}$, see SI~\cite{SI}, Sec. D. 

\subsection{E. Exciton calculations}
We have developed a PyMEX programme, a Python package for Moir\'{e} EXcitons\footnote{The updated package will be freely available on GitHub (\url{https://github.com/imaitygit/PyMEX})}, to solve the Bethe-Salpeter Equation (BSE). The software package uses the mpi4py~\cite{MPI4PY}, numpy~\cite{NUMPY}, scipy~\cite{SCIPY}, cython~\cite{CYTHON}, and h5py~\cite{H5PY} libraries. The Kohn-Sham band gaps of monolayer WS$_{2}$ and WSe$_{2}$ are corrected using the GW method through rigid shifts~\cite{Filipcomputational2015}. We note that by including only a few valence and conduction bands while constructing the BSE Hamiltonian, we do not reach absolute convergence of the lowest lying exciton for small twist angle moir\'{e} unit cell. However, we carefully check the relative convergence of the lowest few excitons with respect to the number of bands. See SI, Sec. E~\cite{SI} for more details of these calculations. Throughout the paper, the onsite interaction for the Keldysh potential is regularized as $W_{0} = U\cdot W_{r=a}$ where $a$ is the pristine unit cell lattice constant of the WS$_{2}$ or WSe$_{2}$, and the parameter $U$ is chosen to be 1~\cite{Fengchengexciton2015}. Furthermore, in all the results presented in the main text, we do not include bare Coulomb interactions. The impact of including the bare Coulomb interaction is discussed in the Supplementary Information (SI)~\cite{SI}.

Low-energy intralayer excitons are localized in the WSe$_2$ layer since it has a smaller band gap than the WS$_2$ layer. Moreover, the low-energy valence and conduction bands of the WSe$_2$ in the twisted bilayer are derived from the $K$ and $K'$ valleys of the flat monolayer. The wavefunctions associated with these valleys are predominantly composed of d-orbitals of W atoms which are only weakly affected by interlayer hybridization (in contrast to states derived from the flat monolayer $\Gamma$ valley)~\cite{Naikintralayer2022, Rossiphason2023, Liimaging2021}. To study intralayer excitons in the twisted heterobilayer, we remove the WS$_2$ layer after the relaxation and carry out large-scale ab initio DFT and BSE calculations on the WSe$_2$ layer with the same atomic structure as in a twisted bilayer system, see Sec.~A of the SI~\cite{SI} for additional details. For the intralayer excitons at twist angles of 16.1° and 5.7°, we use a 15×15×1 grid and a 9×9×1 grid, respectively. A $5\times5\times1$ grid is used for a twist angle of $4.7^\circ$, and a $3\times3\times1$ grid is used for all other twisted bilayer calculations when setting up the BSE Hamiltonian. Results of convergence tests are shown in Sec.~E of the SI~\cite{SI}. 

For the interlayer excitons, we have used a $3\times3\times1$ $k$-grid, 12 valence, and 28 conduction bands to construct the BSE Hamiltonian for 5.7$^\circ$ twisted WS$_{2}$/WSe$_{2}$ heterobilayer. We have incorporated spin-orbit coupling effects perturbatively. We also compute the optical conductivity of isolated WSe$_{2}$ and the WS$_{2}$/WSe$_{2}$ heterobilayer at a 5.7$^\circ$ twist angle using the same \textit{k}-grid for comparison, see SI, Sec.~G.

\section{Data availability}
All data used to produce the plots in this paper are available from the corresponding authors upon request. 

\section{Code availability}
The twisted heterobilayer structure construction, atomic relaxations and electronic structure calculations presented in this paper were carried out using publicly available codes. The python package for moiré exciton calculation is available publicly on GitHub (\url{https://github.com/imaitygit/PyMEX}). Our findings can be fully reproduced by the use of these codes and by following the procedure outlined in the paper.  

\section{Acknowledgements}
This project received funding from the European Union’s Horizon 2020 research and innovation program under the Marie Skłodowska-Curie Grant agreement No. 101028468. The authors acknowledge support from the Thomas Young Centre under Grant No. TYC-101. This work used the ARCHER2 UK National Supercomputing Service via our membership of the UK's HEC Materials Chemistry Consortium, which is funded by EPSRC (EP/X035859), and the UK Car-Parrinello Consortium, which is funded by the EPSRC grant EP/X035891/1 and resources provided by the Cambridge Service for Data-Driven Discovery (CSD3) operated by the University of Cambridge Research Computing Service, provided by Dell EMC and Intel using Tier-2 funding from the EPSRC and DiRAC funding from the Science and Technology Facilities Council, and Imperial College Research Computing Service. We thank Valerio Vitale, Mit H. Naik, and Felipe H. da Jornada for their helpful discussions. We thank Emma Regan, Chenhao Jin, and Feng Wang for providing the experimental data.

\section{Author contributions}
J.L. conceived the project. I.M. developed the methods, implemented the algorithms, and performed the calculations. A.A.M. and J.L. supervised the project. All authors contributed to drafting the manuscript and responding to reviewers. 

\section{Competing Interests}
The authors declare no competing financial or non-financial interests.

\section{Figures}

\clearpage
\newpage 

\section{A: Atomic structures}

\begin{table}[h!]
\centering
%\begin{tabular}{|p{2.5cm}|p{2cm}|p{2cm}|p{2cm}|}
\begin{tabular}{|c|c|c|c|}
\hline
Material & Twist angle & Atoms & Moiré length (in \AA) \\
\hline
WS$_{2}$/WSe$_{2}$ (WSe$_{2}$) & 16.1$^\circ$ & 39 (36) & 11.5 \\
\hline
WS$_{2}$/WSe$_{2}$ (WSe$_{2}$) & 5.68$^\circ$ & 525 (273) & 30.4 \\
\hline
WS$_{2}$/WSe$_{2}$ (WSe$_{2}$) & 4.7$^\circ$ & 696 (333) & 34.9 \\
\hline
WS$_{2}$/WSe$_{2}$ (WSe$_{2}$) & 2.8$^\circ$ & 1362 (651) & 48.9 \\
\hline
WS$_{2}$/WSe$_{2}$ (WSe$_{2}$) & 2.2$^\circ$ & 1776 (849) & 55.9 \\
\hline
WS$_{2}$/WSe$_{2}$ (WSe$_{2}$) & 1.3$^\circ$ & 2490 (1191) & 66.2 \\
\hline
WS$_{2}$/WSe$_{2}$ (WSe$_{2}$) & 0.4$^\circ$ & 3111 (1488) & 74 \\
\hline
WS$_{2}$/WSe$_{2}$ (WSe$_{2}$) & 0$^\circ$ & 2775 (1323) & 69.7 \\
\hline
\end{tabular}
\caption{Summary of the heterobilayer and the twist angles considered in this
paper for all the atomic, electronic, and excitonic calculations. For all the
intralayer electron and exciton calculations, we isolated WSe$_{2}$ from the
relaxed heterobilayer as mentioned in the main text. The unit-cell lattice
constants for the WSe$_{2}$ and WS$_{2}$ layers were set to 3.32 and 3.18
\AA~while generating the moiré patterns. However, to reduce computational
costs, we simulated the 0$^\circ$ twist angle with the unit-cell lattice
constants for the WSe$_{2}$ and WS$_{2}$ layers set to 3.32 and 3.169 \AA,
respectively.}
\end{table}

\section{B: Theoretical Formulations}
\subsection{I: Construction of the BSE Hamiltonian}
The single-particle electronic wave functions obtained from density functional theory (DFT) are expressed using Wannier functions,

\begin{equation}
\psi_{m{\bf k}}({\bf r}) = \frac{1}{\sqrt{N}} \sum_{{\bf R}} e^{i{\bf k}.{\bf R}} \sum_{n} C^{{\bf k}}_{nm} \phi_{n} ({\bf r}-{\bf R} - {\bf t}_{n})
\label{lcao}
\end{equation}
where  ${\bf R}$ denote the lattice vectors commensurate with the ${\bf k}$-grid used, $C^{\bf k}_{nm}$ the expansion coefficients, ${\bf t}_{n}$ represents the location of the $n$-th Wannier function $\phi_{n}$ at ${\bf R}=0$, and $N$ is the number of unit cells. We obtain the $C^{\bf k}_{nm}$ using the WANNIER90 package~\cite{Pizziwannier902020}. 

%For a unit cell calculation, in an optical excitation process, ${\bf Q}$ is the momentum of the photon that is absorbed by the two-particle state. The magnitude of the photon momentum is usually very small and is thus unimportant in the present context. Its direction, on the other hand, is significant for the nonanalytical long-range exchange term of the BSE. In the case of the degenerate valence bands in cubic crystals, e.g., it leads to the splitting of the excitons into transverse and longitudinal modes. Nonzero ${\bf Q}$ vectors are relevant for bound exciton states in materials with an indirect fundamental gap. Note that all the electronic momentum is within the 1st Brillouin zone. Nonetheless, in the case of supercell calculations, the implication of optical absorption is not immediately evident for ${\bf Q}=0$ due to the presence of zone-folding of the electronic structure. 

The matrix required to set up the Bethe-Salpeter-Equation (BSE) in the Tamm-Dancoff approximation is as follows,
\begin{equation}
\langle cv{\bf k Q}| \hat{H}_{\text{e-h}} |c^{\prime}v^{\prime}{\bf k^{\prime} Q}\rangle  = (\epsilon_{c{\bf k+Q}} - \epsilon_{v{\bf k}}) \delta_{cc^\prime} \delta_{vv^\prime} \delta_{\bf kk^\prime} - D_{cv{\bf k Q}, c^{\prime}v^{\prime}{\bf k^{\prime} Q}} + X_{cv{\bf k Q}, c^{\prime}v^{\prime}{\bf k^{\prime} Q}}
\end{equation}

The direct term is written as~\cite{Rohlfingelectron2000},
\begin{equation}
    D_{cv{\bf k Q}, c^{\prime}v^{\prime}{\bf k^{\prime} Q}} = \int d{\bf x} d{\bf x^\prime} \psi^{*}_{c,\bf{k+Q}}({\bf x}) \psi_{c^\prime,\bf{k^\prime+Q}}({\bf x}) W({\bf r, r^\prime}) \psi_{v{\bf k}} ({\bf x^\prime}) \psi^{*}_{v^\prime{\bf k^\prime}} ({\bf x^\prime})
\label{direct}
\end{equation}
The product of the first two wave functions can be written after replacing ${\bf x}={\bf r}$ and using Eqn. \ref{lcao}. It should be noted that in Ref.~\cite{Rohlfingelectron2000}, ${\bf x}$ is represented as ${\bf r, \sigma}, t$. 

\begin{equation}
\begin{split}
& \psi^{*}_{c\bf{k+Q}}({\bf x}) \psi_{c^\prime\bf{k^\prime+Q}}({\bf x})  \\
& = \frac{1}{N}\sum_{{\bf R}_{j_1},n_1} e^{-i({\bf k+Q}).{\bf R}_{j_1}} {C^{\bf k+Q}_{n_{1}c}}^* \phi^{*}_{n_{1}} ({\bf r-R}_{j_1}-{\bf t}_{n_1}) \sum_{{\bf R}_{j_2}, n_2} e^{i({\bf k^\prime+Q}).{\bf R}_{j_2}} C^{\bf k^\prime+Q}_{n_{2}c^\prime} \phi_{n_{2}} ({\bf r-R}_{j_2}-{\bf t}_{n_2}) \\
& = \frac{1}{N}\sum_{{\bf R}_{j_1},n_1; {\bf R}_{j_2}, n_2} e^{-i({\bf k+Q}).{\bf R}_{j_1}} {C^{\bf k+Q}_{n_{1}c}}^*  e^{i({\bf k^\prime+Q}).{\bf R}_{j_2}} C^{\bf k^\prime+Q}_{n_{2}c^\prime} \phi^{*}_{n_{1}} ({\bf r-R}_{j_1}-{\bf t}_{n_{1}})\phi_{n_{2}} ({\bf r-R}_{j_2}-{\bf t}_{n_{2}}) \\
\end{split}
\label{prod1}
\end{equation}
${\bf R}_{j_{i}}$ denotes the lattice vectors commensurate with the $k$-grid. On the other hand, $n_{1}$ and ${\bf t}_{n_{1}}$ denote all the orbitals within the unit cell at ${\bf R}_{j_1}=0$ and their positions. By using the orthogonality and the localized nature of the Wannier functions, we obtain
\begin{equation}
\int d{\bf r}\  \phi^{*}_{n_{1}} ({\bf r-R}_{j_1}-{\bf t}_{n_{1}})\phi_{n_{2}} ({\bf r-R}_{j_2}-{\bf t}_{n_{2}}) = \delta_{{n_{1},n_{2}}} \delta_{{\bf R}_{j_1} ,{\bf R}_{j_2}} \int d{\bf r}\  |\phi_{n_1}({\bf r} - {\bf R}_{{j}_{1}} - {\bf t}_{n_1})|^{2}
\label{orthogonal}
\end{equation}
As a result, Eqn.~\ref{prod1} simplies as the following
\begin{equation}
\begin{split}
  & \frac{1}{N}\sum_{{\bf R}_{j_1},n_1} e^{-i({\bf k+Q}).{\bf R}_{j_1}} {C^{\bf k+Q}_{n_{1}c}}^*  e^{i({\bf k^\prime+Q}).{\bf R}_{j_1}} C^{\bf k^\prime+Q}_{n_{1}c^\prime} \times |\phi_{n_1}({\bf r} - {\bf R}_{{j}_{1}} - {\bf t}_{n_1})|^{2}  \\
= & \frac{1}{N}\sum_{{\bf R}_{j_1},n_1} e^{-i({\bf k- k^\prime}).{\bf R}_{j_1}} {C^{\bf k+Q}_{n_{1}c}}^* C^{\bf k^\prime+Q}_{n_{1}c^\prime} \times |\phi_{n_1}({\bf r} - {\bf R}_{{j}_{1}} - {\bf t}_{n_1})|^{2}
\end{split}
\end{equation}

Similarly, the product of the last two wave functions in Eqn.~\ref{direct} is written as,
\begin{equation}
\begin{split}
& \psi_{v\bf{k}}({\bf r^\prime}) \psi^{*}_{v^\prime\bf{k^\prime}}({\bf r^\prime})  \\
& = \frac{1}{N}\sum_{{\bf R}_{j_3},n_3} e^{i{\bf k}.{\bf R}_{j_3}} {C^{\bf k}_{n_{3}v}} \phi_{n_{3}} ({\bf r^\prime -R}_{j_3}-{\bf t}_{n_3}) \sum_{{\bf R}_{j_4}, n_4} e^{-i{\bf k^\prime}.{\bf R}_{j_4}} {C^{\bf k^\prime}_{n_{4}v^\prime}}^{*} \phi_{n_{4}}^{*} ({\bf r^\prime -R}_{j_4}-{\bf t}_{n_4}) \\
\end{split}
\label{prod2}
\end{equation}
The use of the orthogonality and the localized nature of the Wannier functions leads to simplification
\begin{equation}
\frac{1}{N} \sum_{{\bf R}_{j_3},n_3} e^{i ({\bf k - k^\prime}).{\bf R}_{j_3}}{C^{\bf k}_{n_{3}v}} {C^{\bf k^\prime}_{n_{3}v^\prime}}^* \times |\phi_{n_3}({\bf r^\prime} - {\bf R}_{{j}_{3}} - {\bf t}_{n_3})|^{2}
\end{equation}

As a result, the direct term simplifies to the following
\begin{equation}
\begin{split}
D_{cv{\bf k Q}, c^{\prime}v^{\prime}{\bf k^{\prime} Q}} \\
& =\frac{1}{N^2} \int d{\bf r} d{\bf r^\prime} \sum_{{\bf R}_{j_1},n_{1};{\bf R}_{j_3},n_{3}} e^{-i ({\bf k - k^\prime}).({\bf R}_{j_1}-{\bf R}_{j_3})} W({\bf r, r^\prime}) {C^{\bf k+Q}_{n_{1}c}}^* C^{\bf k^\prime+Q}_{n_{1}c^\prime} {C^{\bf k}_{n_{3}v}} {C^{\bf k^\prime}_{n_{3}v^\prime}}^* \\
& \times |\phi_{n_1}({\bf r} - {\bf R}_{{j}_{1}} - {\bf t}_{n_1})|^{2} |\phi_{n_3}({\bf r^\prime} - {\bf R}_{{j}_{3}} - {\bf t}_{n_3})|^{2} \\ 
\end{split}
\end{equation}
We apply a \textit{translation} operator, so that, ${\bf r} \to ({\bf r} +  {\bf R}_{j_{3}}+{\bf t}_{n_3})$. Also, we replace the Screened Coulomb interaction using Keldysh potential that depends on the difference (${\bf r-r^\prime}$). The Direct term can thus be simplified as
\begin{equation}
\begin{split}
& =\frac{1}{N^2} \int d{\bf r} d{\bf r^\prime} \sum_{{\bf R}_{j_1},n_{1};{\bf R}_{j_3},n_{3}} e^{-i ({\bf k - k^\prime}).({\bf R}_{j_1}-{\bf R}_{j_3})} W({\bf r- r^\prime}) {C^{\bf k+Q}_{n_{1}c}}^* C^{\bf k^\prime+Q}_{n_{1}c^\prime} {C^{\bf k}_{n_{3}v}} {C^{\bf k^\prime}_{n_{3}v^\prime}}^* \\
& \times |\phi_{n_1}({\bf r} + ({\bf R}_{{j}_{3}}-{\bf R}_{{j}_{1}}) + ({\bf t}_{n_3} - {\bf t}_{n_1})|^{2} |\phi_{n_3}({\bf r^\prime})|^{2} 
\end{split}
\end{equation}
As both ${\bf R}_{j_{1}}, {\bf R}_{j_{3}}$ form periodic supercells, and the equation above only depends on the difference between the two, we can perform the double sum with ${\bf R}_{j_{3}} - {\bf R}_{j_{1}} = {\bf R}$. Therefore, for every ${\bf R}_{j_{1}} = {\bf R}_{j_{3}}$, we will have 1 (for the exponential and $\phi_{n_{1}}$ term). So, the equation further simplifies
\begin{equation}
\begin{split}
& =\frac{1}{N^2} \int d{\bf r} d{\bf r^\prime} \sum_{{\bf R},n_{1};n_{3}} e^{i ({\bf k - k^\prime}).{\bf R}} W({\bf r- r^\prime}) {C^{\bf k+Q}_{n_{1}c}}^* C^{\bf k^\prime+Q}_{n_{1}c^\prime} {C^{\bf k}_{n_{3}v}} {C^{\bf k^\prime}_{n_{3}v^\prime}}^* \\
& \times |\phi_{n_1}({\bf r} + {\bf R} + ({\bf t}_{n_3} - {\bf t}_{n_1})|^{2} |\phi_{n_3}({\bf r^\prime})|^{2} \times N \\
& \approx \frac{1}{N} \sum_{{\bf R},n_{1};n_{3}} {C^{\bf k+Q}_{n_{1}c}}^* C^{\bf k^\prime+Q}_{n_{1}c^\prime} {C^{\bf k}_{n_{3}v}} {C^{\bf k^\prime}_{n_{3}v^\prime}}^* W({\bf R} + ({\bf t}_{n_3} - {\bf t}_{n_1})) e^{i ({\bf k - k^\prime}).{\bf R}}
\end{split}
\end{equation}

{\bf Comment on computational efficiency}: The single-particle expansion of the direct term in Eqn.~\ref{direct} results in eight summations, involving four wave functions, each requiring two summations (over $n_{i}$ and ${\bf R}_{{j}_{i}}$)). Exploitation of the localization and orthogonalization of Wannier functions, along with replacing the screened interaction with the Keldysh potential, substantially simplifies the computation, reducing it to only three summations. 

Similarly, we can simplify the computation of the Exchange term 
\begin{equation}
\frac{1}{N} \sum_{{\bf R},n_{1},n_{3}} e^{i {\bf Q}.{\bf R}} {C^{\bf k+Q}_{n_{1}c}}^* {C^{\bf k}_{n_{1}v}} C^{\bf k^\prime+Q}_{n_{3}c^\prime} {C^{\bf k^\prime}_{n_{3}v}} {C^{\bf k^\prime}_{n_{3}v^\prime}}^* V({\bf R} + ({\bf t}_{n_{3}} - {\bf t}_{n_{1}}))
\end{equation}

The BSE is expressed as (presented in the main text for ${\bf Q} =0$)
\begin{equation}
\begin{split}
& \langle cv{\bf k Q}| \hat{H}_{e-h} |c^{\prime}v^{\prime}{\bf k^{\prime} Q}\rangle \\
& = (\epsilon_{c{\bf k+Q}} - \epsilon_{v{\bf k}}) \delta_{cc^\prime} \delta_{vv^\prime} \delta_{\bf kk^\prime} \\
& -\frac{1}{N} \sum_{{\bf R},n_{1};n_{3}} {C^{\bf k+Q}_{n_{1}c}}^* C^{\bf k^\prime+Q}_{n_{1}c^\prime} {C^{\bf k}_{n_{3}v}} {C^{\bf k^\prime}_{n_{3}v^\prime}}^* W({\bf R} + ({\bf t}_{n_3} - {\bf t}_{n_1})) e^{i ({\bf k - k^\prime}).{\bf R}} \\
& + \frac{1}{N} \sum_{{\bf R},n_{1},n_{3}} {C^{\bf k+Q}_{n_{1}c}}^* {C^{\bf k}_{n_{1}v}} C^{\bf k^\prime+Q}_{n_{3}c^\prime} {C^{\bf k^\prime}_{n_{3}v^\prime}}^* V({\bf R} + ({\bf t}_{n_{3}} - {\bf t}_{n_{1}})) e^{i {\bf Q}.{\bf R}}
\end{split}
\label{bse}
\end{equation}

\subsection{II: Comparison with previous theory}
The BSE in Ref.~\cite{Fengchengexciton2015} has the following form
\begin{equation}
\begin{split}
& \langle cv{\bf k Q}| \hat{H}_{e-h} |c^{\prime}v^{\prime}{\bf k^{\prime} Q}\rangle \\
& = (\epsilon_{c{\bf k+Q}} - \epsilon_{v{\bf k}}) \delta_{cc^\prime} \delta_{vv^\prime} \delta_{\bf kk^\prime} \\
& - ({\mathcal{U}^{\bf k + Q}_{c}}^\dagger \mathcal{U}^{\bf k^\prime + Q}_{c^\prime}) ({\mathcal{U}^{\bf k^\prime}_{v^\prime}}^\dagger \mathcal{U}^{\bf k}_{v}) V_{\bf k - k^\prime} + ({\mathcal{U}^{\bf k + Q}_{c}}^\dagger \mathcal{U}^{\bf k}_{v}) ({\mathcal{U}^{\bf k^\prime}_{v^\prime}}^\dagger \mathcal{U}^{\bf k^\prime + Q}_{v}) V_{\bf Q}\\
\end{split}
\label{bsewu}
\end{equation}
$V_{\bf k} = \sum_{\bf R} e^{i{\bf k. R}} V({\bf R})$. We examine the equivalence of our Hamiltonian for the Bethe-Salpeter equation, as presented in Equation~\ref{bse}. The direct term in Eqn.\ref{bse} simplifies if ${\bf t}_{n_{3}}-{\bf t}_{n_{1}} = 0$:
\begin{equation}
\begin{split}
D & = \sum_{n_{1},n_{3}} {C^{\bf k+Q}_{n_{1}c}}^* C^{\bf k^\prime+Q}_{n_{1}c^\prime} {C^{\bf k}_{n_{3}v}} {C^{\bf k^\prime}_{n_{3}v^\prime}}^* \times \frac{1}{N} \sum_{\bf R} W({\bf R}) e^{i ({\bf k - k^\prime}).{\bf R}} \\
& = \sum_{n_{1}} {C^{\bf k+Q}_{n_{1}c}}^* C^{\bf k^\prime+Q}_{n_{1}c^\prime} \times \sum_{n_{3}} {C^{\bf k}_{n_{3}v}} {C^{\bf k^\prime}_{n_{3}v^\prime}}^*  \times V_{\bf k - k^\prime} \\
& = ({\mathcal{U}^{\bf k + Q}_{c}}^\dagger \mathcal{U}^{\bf k^\prime + Q}_{c^\prime}) ({\mathcal{U}^{\bf k^\prime}_{v^\prime}}^\dagger \mathcal{U}^{\bf k}_{v}) V_{\bf k - k^\prime}
\end{split}
\end{equation}
Note that, while making the identification with Eq.~\ref{bsewu}, we have summed over the basis orbitals. Very similar to the direct term, the exchange term can also be recovered.

\section{C: Technical details}
\subsection{I: Obtaining the Coefficients}
We obtain the coefficients from our Wannier functions using the unitary matrix, $C^{\bf k}_{nm} = {U^{\bf k}_{mn}}^\dagger$. We do not construct the maximally localized Wannier functions, as they do not change the solutions of the BSE significantly (see Sec.~E). More details on the construction of the Wannier basis are provided in the main text.

\subsection{II. Quality of Wannier functions for intralayer excitons}
\begin{figure}[ht!]
    \centering
    \includegraphics[scale=0.35]{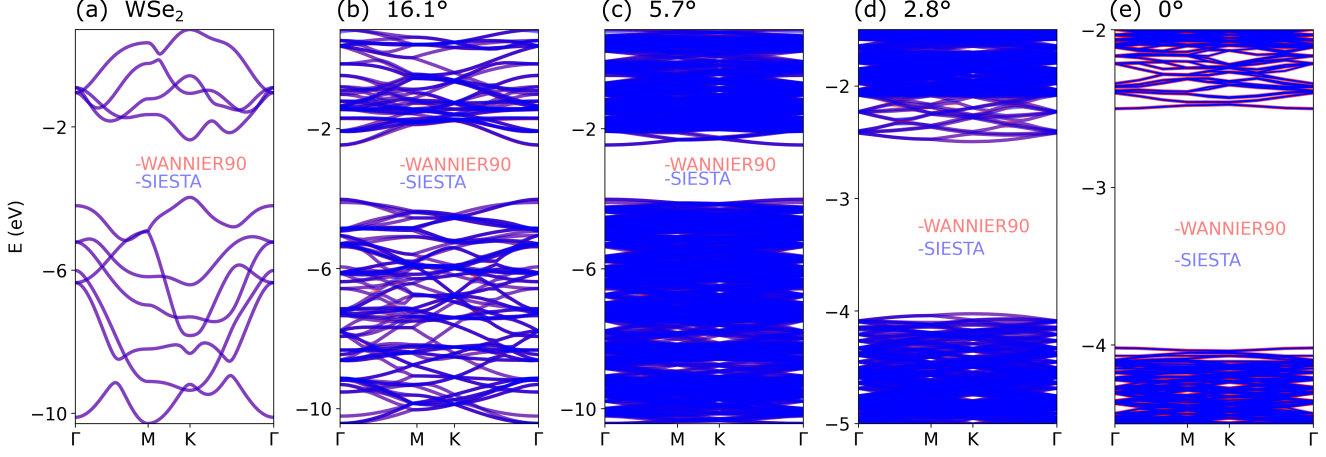}
    \caption{Comparison of the band structures (for intralayer exciton) computed using Wannier90 and SIESTA along the high-symmetry paths. For small twist angles, only a small window near the band gaps is shown for clarity.}
\end{figure}
\subsection{III. Quality of Wannier functions for interlayer excitons}
\begin{figure}[ht!]
    \centering
    \includegraphics[scale=0.35]{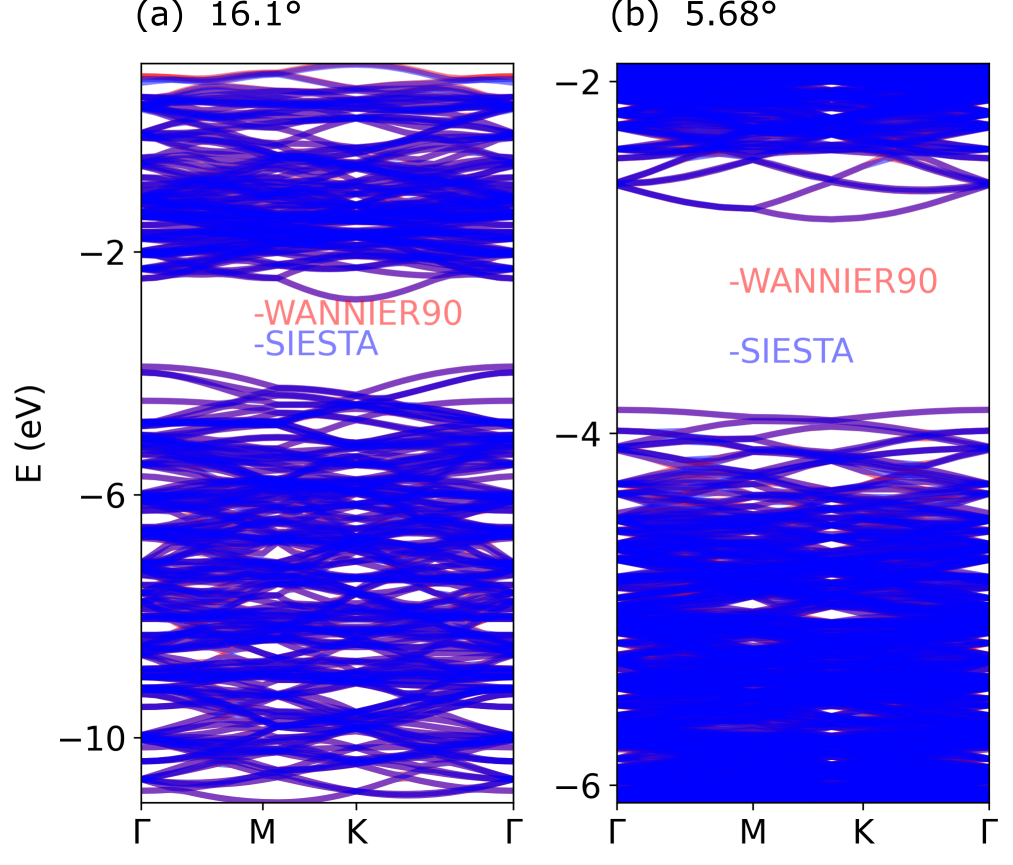}
    \caption{Comparison of the band structures (for interlayer exciton) computed using Wannier90 and SIESTA along the high-symmetry paths. For small twist angles, only a small window near the band gaps is shown for clarity.}
\end{figure}

\subsection{IV: Fixing gauge for the coefficients}
We fix the phase by choosing the sum of the basis set coefficients to be real~\cite{Rohlfingelectron2000}.

\subsection{V: Keldysh potential}
We briefly show how the Keldysh potential is computed from a classical electrostatic problem. A test charge located within a layer $j$ in a dielectric medium obeys the Gauss law.
\begin{equation}
\epsilon_{0} \nabla.{\bf E} = \rho_{f} + \rho_{b} 
\label{gaussinmedium}
\end{equation}
Given our specific focus on 2D materials, the induced bound and free charges become localized within a layer, and there are no charges present in the vacuum. Therefore, the free charge can be written as $\rho_{f} = \rho^{j}({\bf r}) \delta(z-d_{j})$. On the other hand, the bound charge can be induced in any layer and depends on the polarizabilities of individual layers. Therefore, $\rho_{b} = -\sum_{i} \alpha_{i} (\nabla_{\parallel}.{\bf E})_{i} \delta(z-d_{i})$, where $i$ can be any layer in the multilayer system, and $\alpha$ is dielectric constant. The divergence operator contains only in-plane components. The Eqn.~\ref{gaussinmedium} becomes
\begin{equation}
-\epsilon_{0} \nabla^{2}\phi({\bf r}) = \rho^{j}({\bf r}) \delta(z-d_{j}) + \sum_{i} \alpha_{i} (\nabla^{2}_{\parallel} \phi({\bf r}))_{i} \delta(z-d_{i})
\label{GM2}
\end{equation}
We perform a Fourier transform (FT) on the above equation, and the left-hand side becomes
\begin{equation}
\begin{split}
& -\epsilon_{0} \int (\nabla^{2}\phi({\bf r})) e^{-i{\bf k. r}} d{\bf r} \\
& = \epsilon_{0} (k_{\parallel}^{2} + k_{\perp}^{2}) \phi({\bf k_{\parallel}, k_{\perp}})
\end{split}
\end{equation}
In the above, the surface terms vanish, and the FT has $k_{\parallel}, k_{\perp}$. The in-plane and out-of-plane momentum components need different treatments as we deal with 2D materials. We use the following equation, $\nabla.{(f{\bf A})} = f (\nabla . {\bf A}) + \nabla f . {\bf A}$; 
\begin{equation}
(\nabla. \nabla \phi) e^{-i{\bf k.r}} = \nabla.{( e^{-i{\bf k.r}} \nabla \phi)} - (\nabla e^{-i{\bf k.r}}) . \nabla \phi = \nabla.{( e^{-i{\bf k.r}} \nabla \phi)} + \phi (\nabla^{2} e^{-i{\bf k.r}}) - \nabla. (\phi \nabla e^{-i{\bf k.r}}) 
\end{equation}
The Fourier transform (FT) of the first term of the right-hand-side of Eqn.~\ref{GM2} is
\begin{equation}
\begin{split}
& \int \rho^{j}({\bf r}) \delta (z-d_{j}) e^{-i{\bf k}.{\bf r}} d{\bf r} \\
= & \int \rho^{j}({\bf r}_{in},z=d_{j}) e^{-i{\bf k_{\parallel}}.{\bf r}_{in}} e^{-i k_{\perp} d_{j}} d{\bf r}_{in} \\
= & \rho^{j}({\bf k_{\parallel}}) e^{-i k_{\perp} d_{j}}
\end{split}
\end{equation}
In the above equation, we have used a property of the delta function, $\int f(z) \delta (z-d_{j}) dz = f(d_{j})$. The FT of the second term on the right-hand side of Eqn.~\ref{GM2} can be simplified using similar steps as outlined above
\begin{equation}
\begin{split}
& \sum_{i} \alpha_{i} (\nabla^{2}_{\parallel} \phi({\bf r}))_{i} \delta(z-d_{i}) e^{-i{\bf k . r}} d{\bf r} \\
& = - k_{\parallel}^{2}\sum_{i} \alpha_{i} \phi({\bf k_{\parallel}}, d_{i}) e^{-i k_{\perp} d_{i}}
\end{split}
\end{equation}
Putting everything back together,
\begin{equation}
\epsilon_{0} (k_{\parallel}^{2} + k_{\perp}^{2}) \phi({\bf k_{\parallel}}, k_{\perp}) = \rho^{j}({\bf k_{\parallel}}) e^{-i k_{\perp} d_{j}} -  k_{\parallel}^{2}\sum_{i} \alpha_{i} \phi({\bf k_{\parallel}}, d_{i}) e^{-i k_{\perp} d_{i}}
\label{GM3}
\end{equation}

We want the effective potential without $k_{\perp}$ and therefore, we use the Fourier inversion theorem. One very prominent Fourier transformation is : $\mathcal{FT} (e^{-a|x|}) = \frac{2a}{s^{2} + a^{2}}$. 
 Note: A very nice trick is that in the limit of $a\to 0$ one finds that $\frac{2a}{s^{2} + a^{2}} \to \delta(s)$ and $e^{-a|x|} \to 0$. Another noteworthy relation is (not necessary in the current calculation) $\mathcal{FT} (e^{-a^{2}x^{2}}) = \frac{\sqrt{\pi}}{a} (e^{-s^{2}/4a^{2}})$. Again the limiting case is interesting.
The Eqn.~\ref{GM3} can be expressed as
\begin{equation}
\begin{split}
& \epsilon_{0} (k_{\parallel}^{2} + k_{\perp}^{2}) \phi({\bf k_{\parallel}}, k_{\perp}) = \rho^{j}({\bf k_{\parallel}}) e^{-i k_{\perp} d_{j}} -  k_{\parallel}^{2}\sum_{i} \alpha_{i} \phi({\bf k_{\parallel}}, d_{i}) e^{-i k_{\perp} d_{i}} \\
& 2{k_{\parallel}} \epsilon_{0} \phi({\bf k_{\parallel}}, k_{\perp}) = 2{k_{\parallel}} \frac{\rho^{j}({\bf k_{\parallel}}) e^{-i k_{\perp} d_{j}}}{(k_{\parallel}^{2} +  k_{\perp}^{2})} - 2{k_{\parallel}} \frac{  k_{\parallel}^{2}\sum_{i} \alpha_{i} \phi({\bf k_{\parallel}}, d_{i}) e^{-i k_{\perp} d_{i}}}{(k_{\parallel}^{2} + k_{\perp}^{2})}
\end{split}
\label{GM4}
\end{equation}

The Inverse FT of the left-hand-side becomes
\begin{equation}
\int 2 k_{\parallel} \epsilon_{0} \phi({\bf k_{\parallel}}, k_{\perp}) e^{ik_{\perp}z} dk_{\perp} = 2 k_{\parallel} \epsilon_{0} \phi({\bf k_{\parallel}},z)
\end{equation}
The Inverse FT of the first term in the right-hand-side of Eqn.~\ref{GM4} becomes (Note the use of aforementioned Fourier inverse transform)
\begin{equation}
\begin{split}
& \int 2{k_{\parallel}} \frac{\rho^{j}({\bf k_{\parallel}}) e^{-i k_{\perp} d_{j}}}{(k_{\parallel}^{2} +  k_{\perp}^{2})} e^{ik_{\perp}z} dk_{\perp} \\
= & \rho^{j}({\bf k_{\parallel}}) \int \frac{2k_{\parallel}}{(k_{\parallel}^{2} +  k_{\perp}^{2})} e^{ik_{\perp}(z-d_{j})} \\
= & \rho^{j}({\bf k_{\parallel}}) e^{-k_{\parallel} |z-d_{j}|}
\end{split}
\end{equation}
Very similar analysis of the last term and everything put together we have,
\begin{equation}
2 k_{\parallel} \epsilon_{0} \phi({\bf k_{\parallel}},z) = \rho^{j}({\bf k_{\parallel}}) e^{-k_{\parallel} |z-d_{j}|} - k_{\parallel}^{2} \sum_{i} \alpha_{i} \phi({\bf k_{\parallel}}, d_{i}) e^{-k_{\parallel} |z - d_{i}|}
\end{equation}
When evaluated at different layers (as the inter-layer separation is primarily vacuum),
\begin{equation}
\begin{split}
& 2 k_{\parallel} \epsilon_{0} \phi({\bf k_{\parallel}},d_{l}) = \rho^{j}({\bf k_{\parallel}}) e^{-k_{\parallel} |d_{l}-d_{j}|} - k_{\parallel}^{2} \sum_{i} \alpha_{i} \phi({\bf k_{\parallel}}, d_{i}) e^{-k_{\parallel} |d_{l} - d_{i}|}\\
& \sum_{i} M_{li} \phi_{i}({\bf k_{\parallel}}) = \rho_{l}^{j}({\bf k_{\parallel}})
\end{split}
\label{gausslaw}
\end{equation}
In the above, $M_{li} = 2 k_{\parallel}\epsilon_{0} + k_{\parallel}^{2} \alpha_{l}$ when $l=i$, and $M_{li} = (k_{\parallel})^{2}\alpha_{i} e^{-k_{\parallel} |d_{l} - d_{i}|}$ when $l \neq i$. Also $\rho_{l}^{j}({\bf k_{\parallel}}) = \rho^{j}({\bf k_{\parallel}}) e^{-k_{\parallel} |d_{l} - d_{i}|}$, and $\phi_{i}({\bf k_{\parallel}}) = \phi({\bf k_{\parallel}}, d_{i})$. 

To calculate the screened Coulomb interaction, we follow the work of Danovich and co-workers~\cite{Danovichlocalized2018}. The authors used classical electrostatics to calculate the potential created by a point charge in a bilayer system where each of the layers is described as a two-dimensional polarizable sheet. The potential in layer $i$ created by a point charge in the same layer is given by $\mathcal{V}_{i}(\mathbf{q})$ and the potential in one layer created by a point charge in the other layer is $\mathcal{W}(\mathbf{q})$. The Fourier transforms of these interactions are found to be (using the same definitions as in the main text)
\begin{equation}
\begin{split}
& \mathcal{V}_{1}(q) = \frac{2\pi (1+ r_{2}q - r_{2}qe^{-2qd})}{[\epsilon_{bg} q (1+r_{1}q)(1+r_{2}q) - r_{1}r_{2}q^{2}e^{-2qd}]}, \\
& \mathcal{V}_{2}(q) = \frac{2\pi (1+ r_{1}q - r_{1}qe^{-2qd})}{[\epsilon_{bg} q (1+r_{1}q)(1+r_{2}q) - r_{1}r_{2}q^{2}e^{-2qd}]}, \\
& \mathcal{W}(q) = \frac{2\pi e^{-qd}}{[\epsilon_{bg} q (1+r_{1}q)(1+r_{2}q) - r_{1}r_{2}q^{2}e^{-2qd}]}. \\
\end{split}
\end{equation}

As a consequence of the large exciton radii in monolayer TMDs and their heterostructures, we focus on the small momentum limit of these interactions given by
\begin{equation}
\begin{split}
& \mathcal{V}_{1}^{\text{small}}(q) = \frac{2\pi}{[\epsilon_{\mathrm{bg}} q (1+q(r_{1} +r_{2}))]}, \\
& \mathcal{V}_{2}^{\text{small}}(q) = \frac{2\pi}{[\epsilon_{\mathrm{bg}} q (1+q(r_{1} +r_{2}))]}, \\
& \mathcal{W}^{\text{small}}(q) = \frac{2\pi}{[\epsilon_{\mathrm{bg}} q (1+q(r_{1} +r_{2} + d))]}.
\end{split}
\end{equation}
These expressions are the Fourier transforms of the standard Keldysh potential. As mentioned in the original manuscript, the interlayer interaction is the same as the intralayer interaction with the only difference that $r_1 + r_2$ is replaced by $r_1 + r_2 + d$.

\clearpage
\newpage

\subsection{VI: Spin-orbit coupling as a perturbation}
We solve the BSE without spin-orbit coupling and include the spin-orbit coupling as a perturbation, as shown in monolayer MoS$_{2}$~\cite{Dianascreening2016}. This approach is computationally much cheaper and can be written as,
\begin{equation}
    \operatorname{Re}[\sigma_{xx}(\omega)] = \frac{\pi e^{2}}{\hbar \omega V m^{2}} \sum_{S,s} |\langle \Omega|\hat{p}_{x}|S\rangle ^{2} \delta(\omega - \omega^{S}_{s}) 
\end{equation}
where $s$ denotes spin, and $\omega_{s}^{S}=\sum\limits_{vc{\bf k}} |A_{vc{\bf k}}^{S}|^{2} \Delta_{vc{\bf k}s}^{SOC}$ is the spin-orbit corrected BSE eigenvalues. The spin-orbit correction required to modify the BSE eigenvalues can be calculated as, $\Delta_{vc{\bf k}s}^{SOC} = (\epsilon_{c{\bf k}} - \epsilon_{v {\bf k}}) + (\Delta \epsilon^{SOC}_{c{\bf k}s}-\Delta \epsilon^{SOC}_{v{\bf k}s})$. For a monolayer of transition metal dichalcogenide calculation, it's straightforward to compute $\Delta_{vc{\bf k}s}^{SOC}$. We have used this approach for all the monolayer calculations presented in this paper. For the twisted bilayer calculations, we explicitly compare the band structure with and without spin-orbit coupling and use the wave functions to extract the spin-orbit coupling for the bands closest to the valence band edge and conduction band edge. The optical conductivity without the spin-orbit coupling can be expressed as~\cite{Ridolfiexcitonic2018},
\begin{equation}
\begin{split}
 \operatorname{Re}[\sigma_{xx}(\omega)] = & \frac{\pi e^{2}}{\hbar \omega V } \sum_{S} |\langle \Omega|\hat{p}_{x}|S\rangle ^{2} \delta(\omega - \omega^{S}) \\
 \propto & \sum_{S} |\langle \Omega|\hat{p}_{x}|S\rangle ^{2} \delta(\omega - \omega^{S}) \\
 \propto &  \sum_{S} |\sum_{cv{\bf k}} A^{S}_{cv{\bf k}} \langle v{\bf k}|p_{x}|c{\bf k}\rangle|^{2} \delta(\omega - \omega^{S})
\end{split}
\end{equation}
While evaluating the optical conductivity, we replaced the delta function with a Gaussian function. Below, we outline the details of computing the momentum operator using our Wannier-derived tight-binding model. 

The momentum matrix elements can be evaluated using ${\bf p}=\frac{i m}{\hbar}[H, {\bf r}]$. We outline steps for evaluating the momentum matrix elements below,
\begin{equation}
\begin{split}
\langle v{\bf k}|{\bf p}|c{\bf k}\rangle \propto & \langle v{\bf k}|i[H, {\bf r}]|c{\bf k}\rangle \\
= & \langle v{\bf k}|i( H{\bf r} - {\bf r} H) |c{\bf k}\rangle \\
= & \frac{1}{N} \sum_{n_{1}n_{2}} \sum_{\bf{RR}^\prime} (C^{\bf k}_{n_{2}v})^{*} C^{\bf k}_{n_{1}c} e^{i{\bf k}\cdot (\bf {R}-\bf {R}^\prime)} \langle n_{2} {\bf R}^\prime | i( H{\bf r} - {\bf r} H) | n_{1}{\bf R}\rangle \\
= & \frac{1}{N} \sum_{n_{1}n_{2}} \sum_{\bf{RR}^\prime} (C^{\bf k}_{n_{2}v})^{*} C^{\bf k}_{n_{1}c} e^{i{\bf k}\cdot (\bf {R}-\bf {R}^\prime)} ( i ({\bf R}-{\bf R}^\prime) + i ({\bf t}_{n_{1}} - {\bf t}_{n_{2}})) \langle n_{2} {\bf R}^\prime | H | n_{1}{\bf R}\rangle \\
= & \sum_{n_{1}n_{2}} (C^{\bf k}_{n_{2}v})^{*} C^{\bf k}_{n_{1}c} \left ( \langle n_{2}{\bf k}|\nabla_{\bf k} H|n_{1} {\bf k}\rangle +  i ({\bf t}_{n_{1}} - {\bf t}_{n_{2}}) \right \langle n_{2}{\bf k}| H|n_{1} {\bf k}\rangle) \\
= & \sum_{n_{1}n_{2}} (C^{\bf k}_{n_{2}v})^{*} C^{\bf k}_{n_{1}c} \langle n_{2}{\bf k}|\nabla_{\bf k} H|n_{1} {\bf k}\rangle + i (E_{v {\bf k}} - E_{c {\bf k}})\sum_{n_{1}} (C^{\bf k}_{n_{1}v})^{*} C^{\bf k}_{n_{1}c} {\bf t}_{n_{1}}
\end{split}
\end{equation}

In the above, we use a complete set of localized basis orbitals to compute
\begin{equation}
\begin{split}
& \langle n_{2} {\bf R}^\prime | ( H{\bf r}) | n_{1}{\bf R}\rangle\\
& =  \langle n_{2} {\bf R}^\prime | H \sum_{{\bf R}^{''}n_{3}} |{\bf R}^{''} n_{3} \rangle \langle {\bf R}^{''} n_{3}| {\bf r} |  n_{1}{\bf R}\rangle \\  
& =  \langle n_{2} {\bf R}^\prime | H | n_{1}{\bf R}\rangle ({\bf R} + {\bf t}_{n_{1}}) \\
\end{split}
\end{equation}
and 
\begin{equation}
\begin{split}
& \langle n_{2} {\bf R}^\prime | ( {\bf r} H)| n_{1}{\bf R}\rangle\\
& =  \langle n_{2} {\bf R}^\prime | {\bf r} \sum_{{\bf R}^{''}n_{3}} |{\bf R}^{''} n_{3} \rangle \langle {\bf R}^{''} n_{3}| H |  n_{1}{\bf R}\rangle \\  
& =  \langle n_{2} {\bf R}^\prime | H | n_{1}{\bf R}\rangle ({\bf R}^\prime + {\bf t}_{n_{2}}) \\
\end{split}
\end{equation}

, $\sum_{n_{2}} H^{\bf k}_{n_{2}n_{1}} (C^{\bf k}_{n_{2}v})^{*} = E_{v {\bf k}} (C^{\bf k}_{n_{1}v})^{*}$, and $\sum_{n_{1}} H^{\bf k}_{n_{2}n_{1}} C^{\bf k}_{n_{1}c} = E_{c {\bf k}} C^{\bf k}_{n_{1}c}$. 
\clearpage 
\newpage 

\section{D: Benchmarking against previous reports}
\subsection{I: Monolayer WS$_{2}$}
\begin{figure}[ht!]
    \centering
    \includegraphics{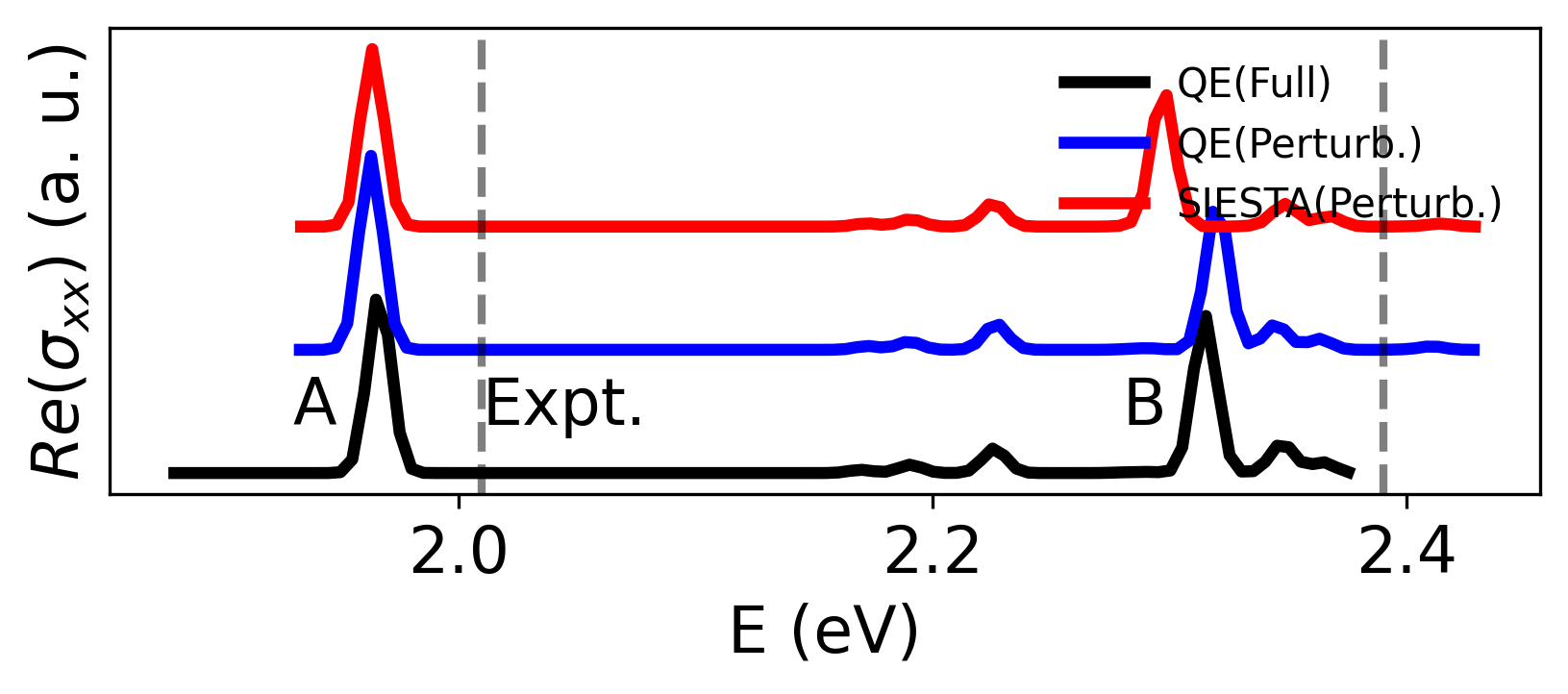}
    \caption{Optical conductivity results of monolayer $\mathrm{WS_{2}}$ with Wannier basis. The GW correction is included as a rigid shift~\cite{Filipcomputational2015}. The experimental results for the A exciton are also marked with dashed lines for samples on a polydimethylsiloxane (PDMS) substrate. The energy can vary depending on the substrates; for example, the A exciton binding energy on a SiO$_{2}$ substrate is 2.01 eV~\cite{Hanickimeasurement2015}, while on a PDMS substrate, it is 2.08 eV~\cite{Niuthickness2018}. We have used a very small smearing of 5 meV with Gaussian functions to represent the $\delta$ functions during the optical conductivity calculations. Calculations with SIESTA/QE (Quantum ESPRESSO) imply the DFT package to generate the single particle wavefunctions to obtain the Wannier functions. We explicitly compute exciton spectra including the spin-orbit-coupling in our DFT (marked as ``Full") and compare them to the perturbative approach (see Sec.~C for details of the perturbative approach). We have used a $39\times 39\times 1$ $k$-grid, 1 valence, and 1 conduction band for the perturbative approach, and 2 valence and 2 conduction bands for the ``full" approach. No substrate effects (i.e. $\epsilon_{r}=1$) are included in our Keldysh potential ~\cite{Berkelbachtheory2013}. The A and B excitons represent the spin-split partners of the transitions at the $K$ point of the Brillouin zone and their separation is a measure of the spin-orbit coupling. The separation between A and B peaks in our calculations is ($\approx$ 340-350 meV) also in good agreement with experimental results ($\approx 370$ meV).}
\end{figure}

\clearpage 
\newpage 
\subsection{II. Monolayer WSe$_{2}$}
\begin{figure}[ht!]
    \centering
    \includegraphics{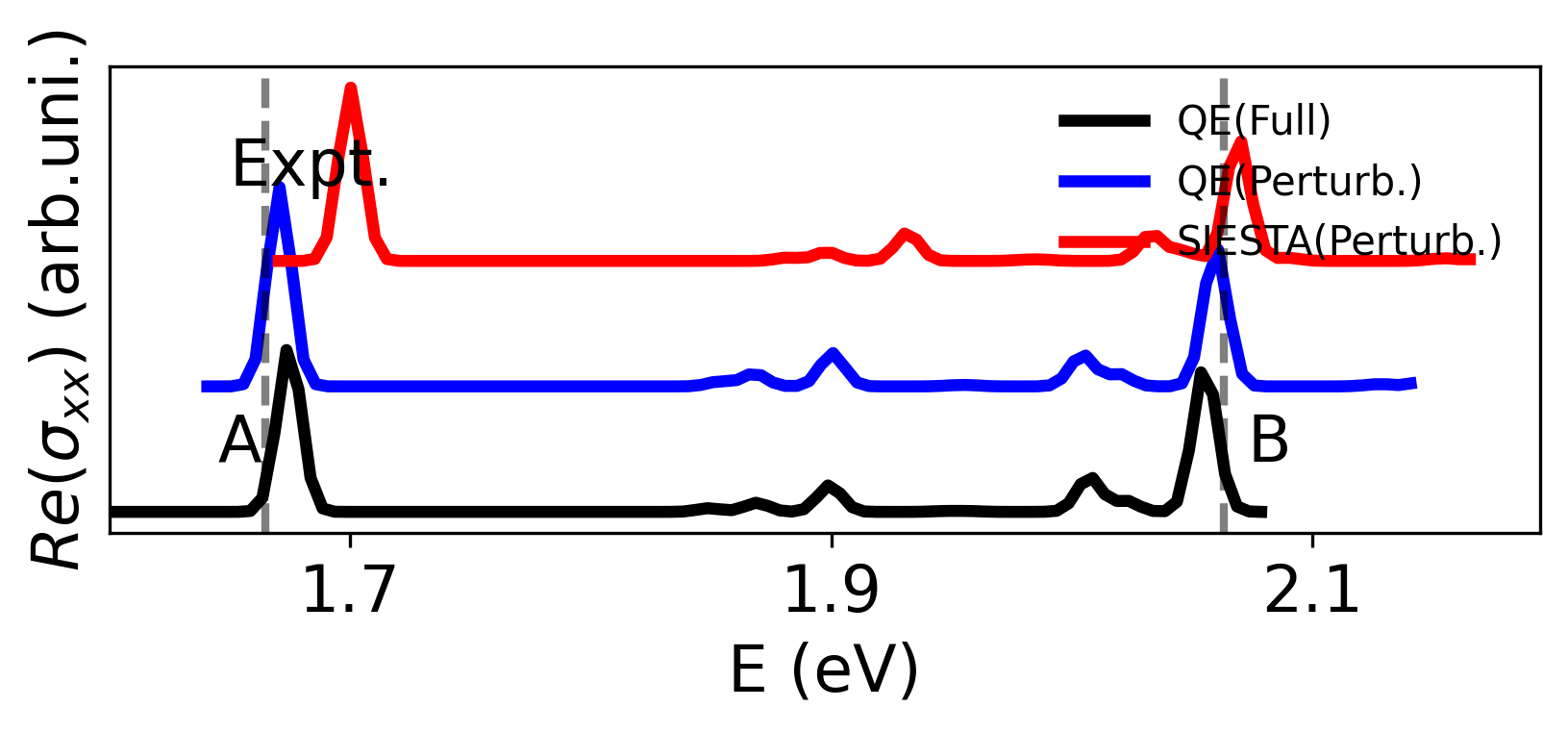}
    \caption{Optical conductivity results of monolayer $\mathrm{WSe_{2}}$ with Wannier basis. The GW correction is included as a rigid shift~\cite{Filipcomputational2015}. The experimental results for the A exciton are also marked with dashed lines for samples on a polydimethylsiloxane (PDMS) substrate. The energy can vary depending on the substrates; for example, the A exciton binding energy on a SiO$_{2}$ substrate is 1.66 eV~\cite{Hanickimeasurement2015}, while on a PDMS substrate, it is 1.74 eV~\cite{Niuthickness2018}. We have used a very small smearing of 5 meV with Gaussian functions to represent the $\delta$ functions during the optical conductivity calculations. Calculations with SIESTA/QE (Quantum Espresso) imply the DFT package to generate the single particle wavefunctions to obtain the Wannier functions. We explicitly compute exciton spectra including the spin-orbit-coupling in our DFT (marked as ``Full") and compare them to the perturbative approach. We have used a $39\times 39\times 1$ $k$-grid, 1 valence, and 1 conduction band for the perturbative approach, and 2 valence and 2 conduction bands for the ``full" approach. No substrate effects are included in our Keldysh potential ~\cite{Berkelbachtheory2013}. The A and B excitons represent the spin-split partners of the transitions at the $K$ point of the Brillouin zone and their separation is a measure of the spin-orbit coupling. The separation between A and B peaks in our calculations is ($\approx$ 370-380 meV) also in good agreement with experimental results ($\approx 398$ meV).}
\end{figure}

\clearpage
\newpage

\subsection{III: Comparison with previous GW-BSE calculations}
\begin{figure}[ht!]
    \centering
    \includegraphics[scale=0.8]{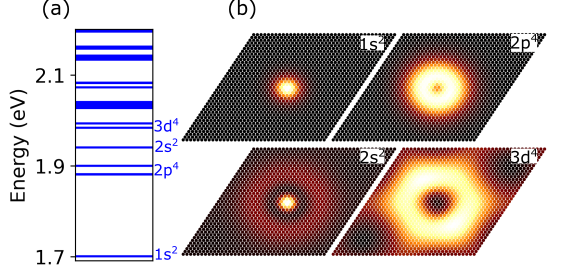}
    \caption{(a) We examine the exciton series of a monolayer of WSe$_{2}$ using our method and observe significant deviations in the BSE eigenvalues from the 2D hydrogen model. Specifically, the degeneracy in energy among states with the same principal quantum number is broken; for instance, $2p^{4}$ is no longer 4-fold degenerate. Furthermore, excitons with the same principal number but higher angular momentum number become lower energy states (e.g., $E_{3d}<E_{3p}<E_{3s}$). These findings are consistent with previous GW-BSE calculations~\cite{Yeprobing2014}. (b) We show corresponding exciton wave functions by fixing the hole at the origin. Each dot represents a pristine WSe$_{2}$ unit cell. In these calculations, we used a $39\times 39\times 1$ $k$-grid, included 2 valence, and 4 conduction bands while constructing the BSE Hamiltonian, and excluded spin-orbit coupling.}
\end{figure}

\clearpage
\newpage

\section{E: Convergence tests}
\subsection{I: One-shot projected Wannier function (WF) vs. maximally localized Wannier function (MLWF)}

\begin{figure}[ht!]
    \centering
    \includegraphics[scale=0.7]{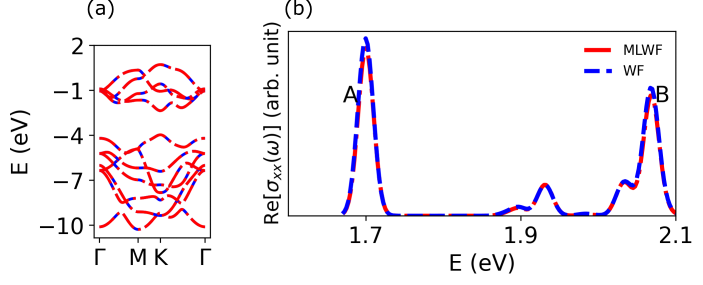}
    \caption{(a),(b): Comparison of electronic band structure and optical conductivity using WF and MLWF as the basis for monolayer WSe$_{2}$. Spin-orbit coupling is included perturbatively, and GW corrections are applied~\cite{Filipcomputational2015}. We employ 2 valence and 4 conduction bands, a $39\times39\times1$ $k$-grid, and a 10 meV smearing to replace the delta functions. Peak A from both calculations is aligned.}
\end{figure}
\clearpage 
\newpage 
\subsection{II: Choice of basis in DFT calculations}
\begin{figure}[ht!]
    \centering
    \includegraphics[scale=0.7]{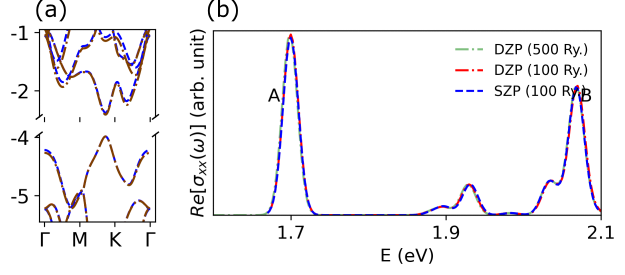}
    \caption{(a),(b): Comparison of electronic band structure and optical conductivity with Wannier basis, where DFT calculations are performed with a single-$\zeta$ plus polarization basis (SZP, with a mesh cutoff of 100 Rydberg) and a double-$\zeta$ plus polarization basis (DZP, with a mesh cutoff of 100 Ry. and 500 Ry.) for monolayer WSe$_{2}$. Spin-orbit coupling is perturbatively included, and GW corrections are included as a rigid shift. We employ 2 valence and 4 conduction bands, a $39\times39\times1$ $k$-grid, and a smearing of 10 meV to replace the delta functions. We align peak A from all the calculations.
    \label{szpvsdzp}
    }
\end{figure}
\clearpage
\newpage

\subsection{III. Convergence of intralayer excitons in twisted $\mathrm{WS_{2}/WSe_{2}}$}

%IM: Done @23/06/2024
\begin{figure}[ht!]
    \centering
    \includegraphics{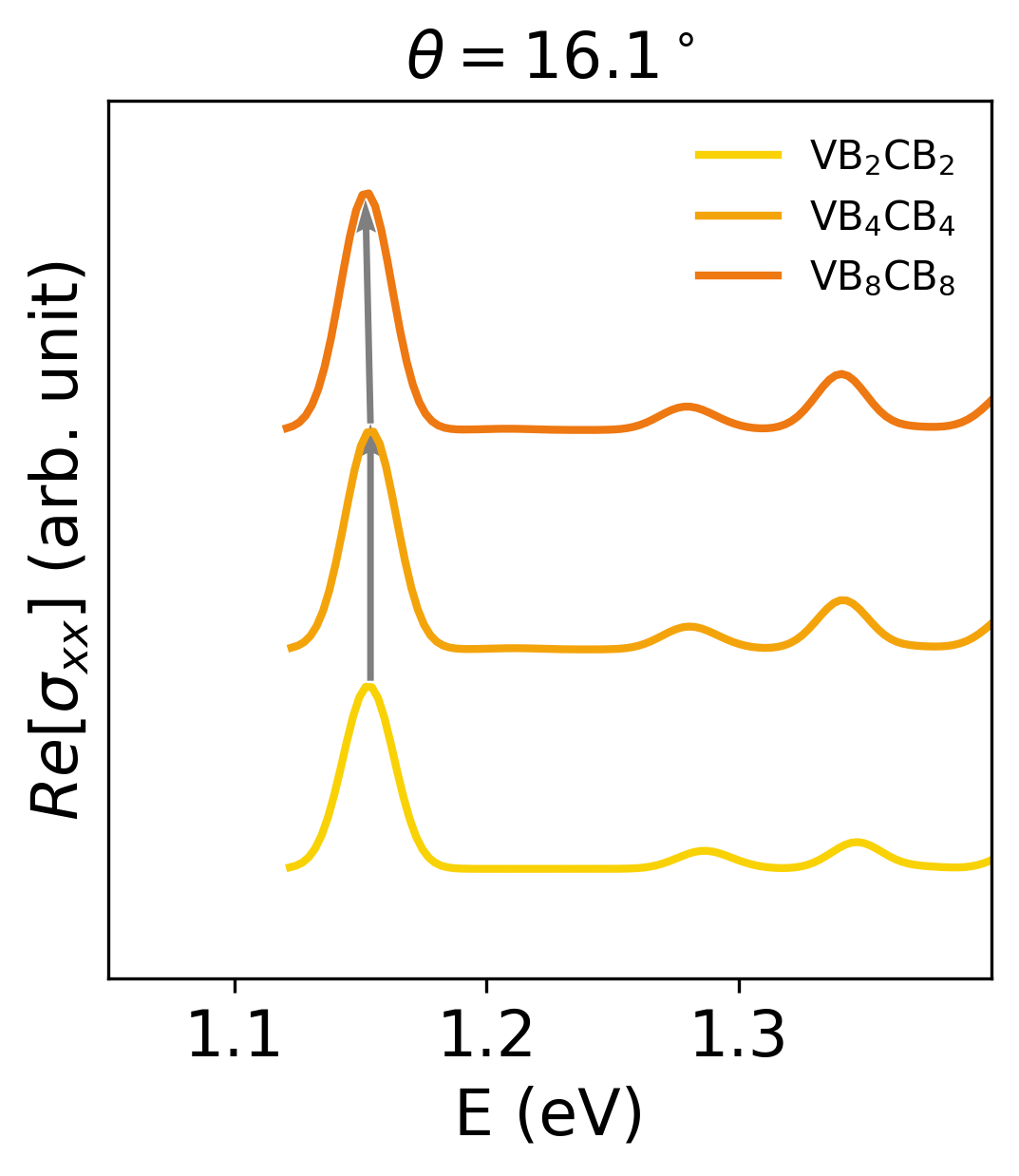}
    \caption{Impact of increasing the number of bands while construction of the BSE Hamiltonian in moir\'{e} systems. The figure shows the results of relaxed WSe$_{2}$ of a 16.1$^\circ$ twisted heterobilayer as representative of intralayer excitons of WSe$_{2}$. We do not include the GW correction and the spin-orbit coupling in the convergence study. We use a $15\times 15\times 1$ k-grid and vary the number of valence bands (VB) and conduction bands (CB). We use arrows as a guide to the eye for the convergence of the A peak.}
\end{figure}

%IM: Done@23/06/2024
\begin{figure}[ht!]
    \centering
    \includegraphics{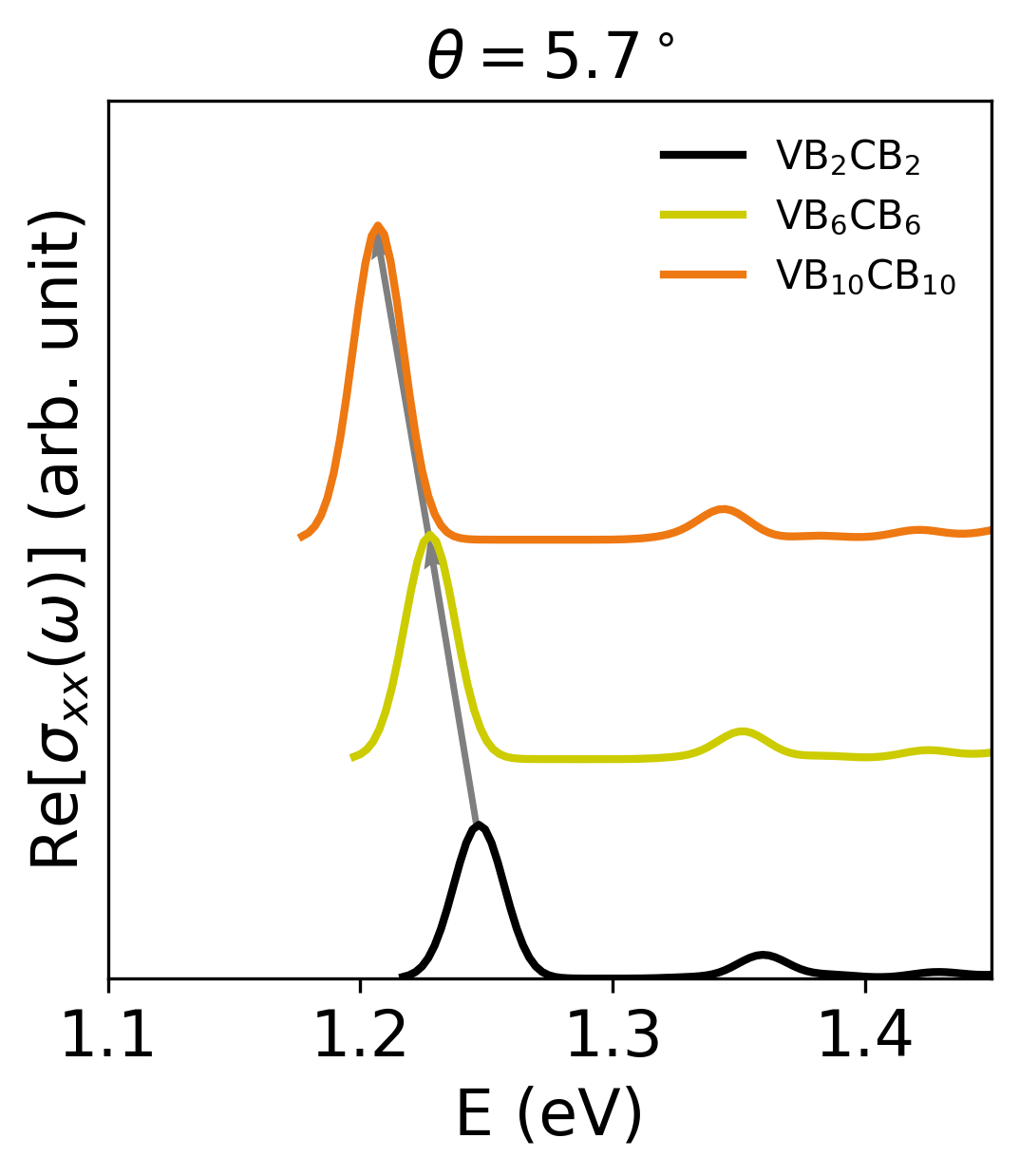}
    \caption{Impact of increasing the number of bands while construction of the BSE Hamiltonian in moir\'{e} systems. The figure shows the results of relaxed WSe$_{2}$ of a 5.7$^\circ$ twisted heterobilayer as representative of intralayer excitons of WSe$_{2}$. We do not include the GW correction and the spin-orbit coupling in the convergence study. We use a $9\times 9\times 1$ k-grid and vary the number of valence bands (VB) and conduction bands (CB). We denote the ``A" peak with dashed lines.}
\end{figure}

%IM: Done@23/06/2024
\begin{figure}[ht!]
    \centering
    \includegraphics{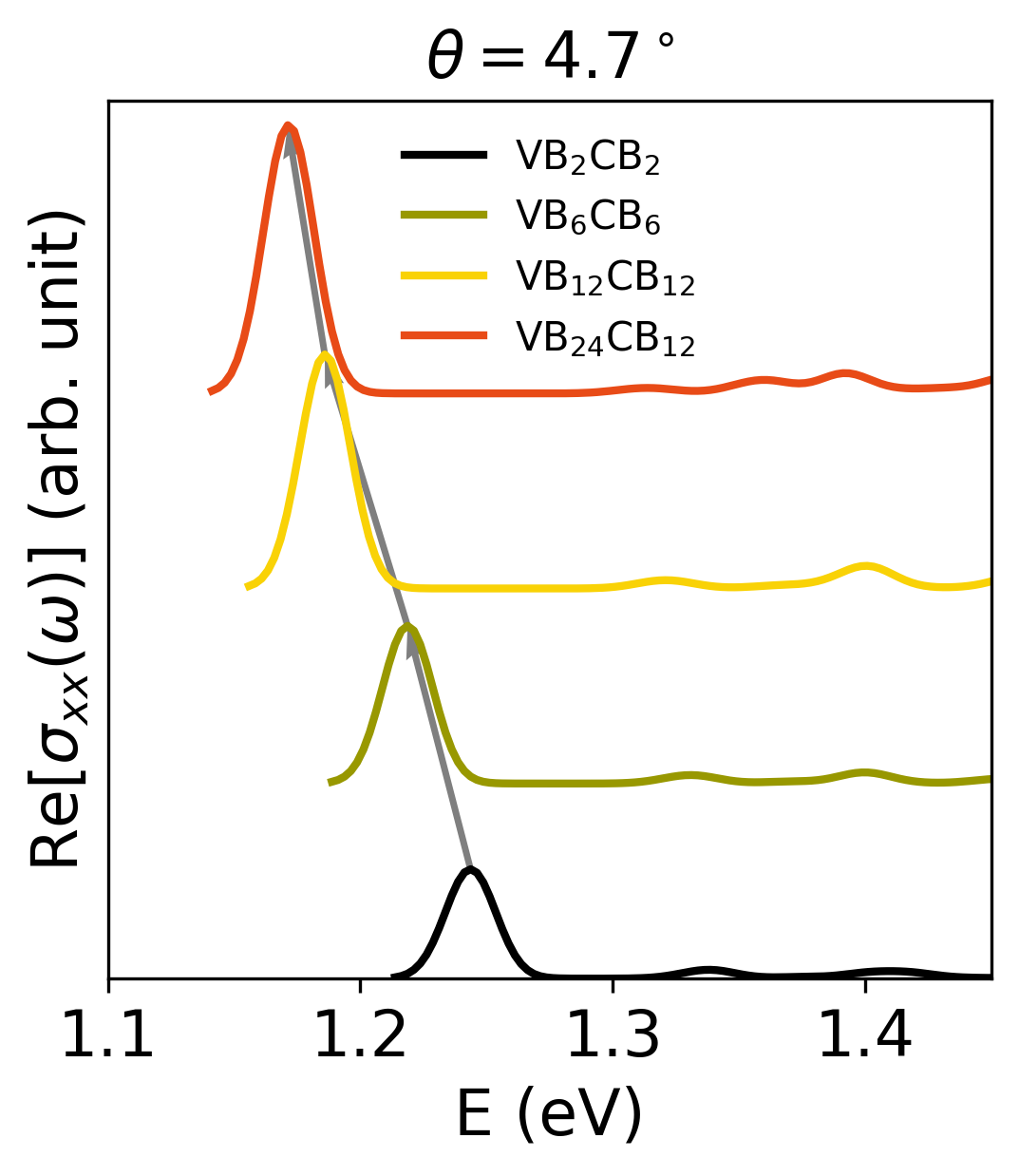}
    \caption{Impact of increasing the number of bands while construction of the BSE Hamiltonian in moir\'{e} systems. The figure shows the results of relaxed WSe$_{2}$ of a 4.7$^\circ$ twisted heterobilayer as representative of intralayer excitons of WSe$_{2}$. We do not include the GW correction and the spin-orbit coupling in the convergence study. We use a $9\times 9\times 1$ k-grid and vary the number of valence bands (VB) and conduction bands (CB). We use arrows as a guide to the eye for the convergence of the A peak.}
\end{figure}

%IM: Done
\begin{figure}[ht!]
    \centering
    \includegraphics{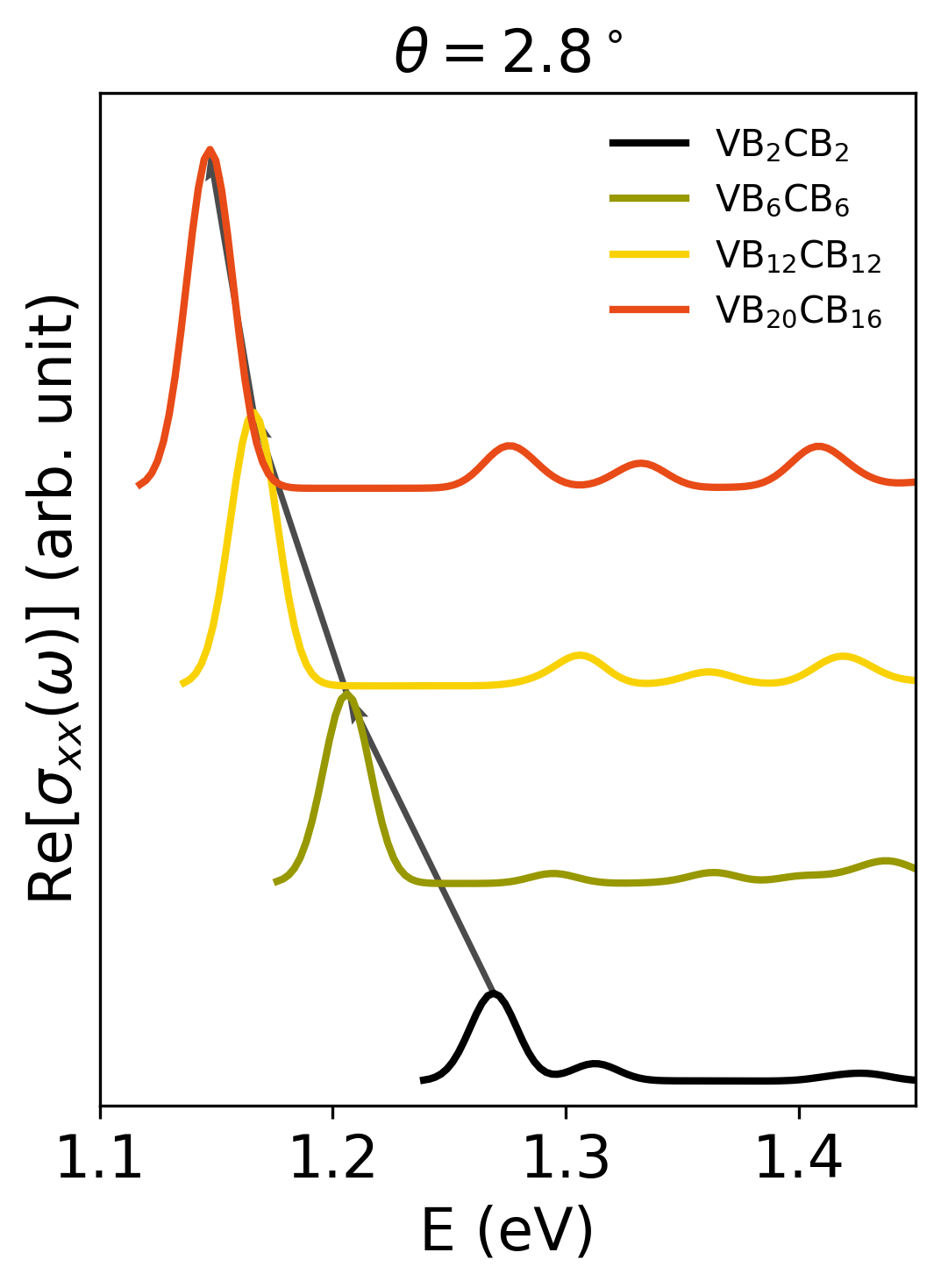}
    \caption{Impact of increasing the number of bands while construction of the BSE Hamiltonian in moir\'{e} systems. The figure shows the results of relaxed WSe$_{2}$ of a 2.8$^\circ$ twisted heterobilayer as representative of intralayer excitons of WSe$_{2}$. We do not include the GW correction and the spin-orbit coupling in the convergence study. We use a $3\times 3\times 1$ k-grid and vary the number of valence bands (VB) and conduction bands (CB). We use arrows as a guide to the eye for the convergence of the A peak.}
\end{figure}

%IM: Done
\begin{figure}[ht!]
    \centering
    \includegraphics{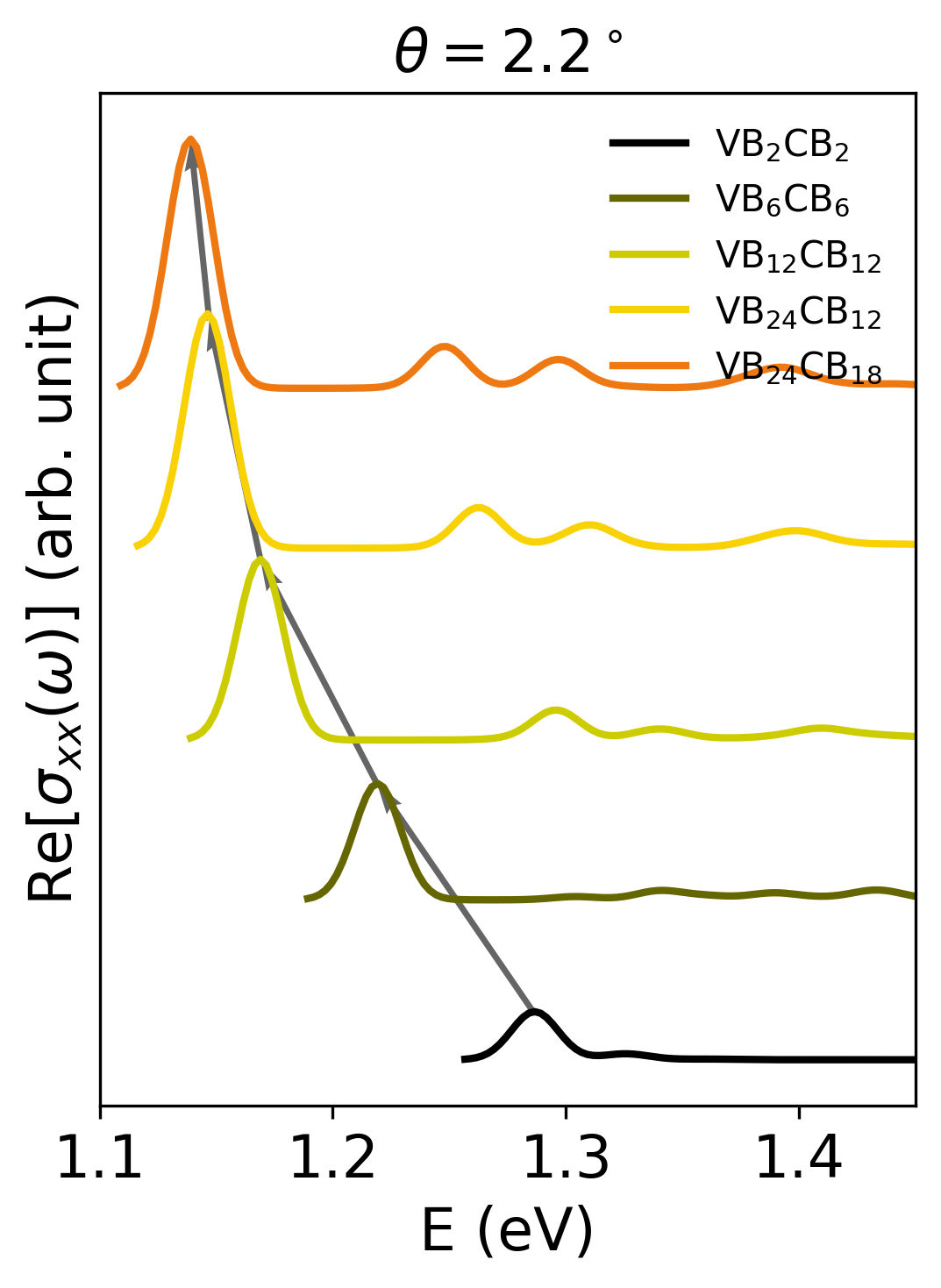}
    \caption{Impact of increasing the number of bands while construction of the BSE Hamiltonian in moir\'{e} systems. The figure shows the results of relaxed WSe$_{2}$ of a 2.2$^\circ$ twisted heterobilayer as representative of intralayer excitons of WSe$_{2}$. We do not include the GW correction and the spin-orbit coupling in the convergence study. We use a $3\times 3\times 1$ k-grid and vary the number of valence bands (VB) and conduction bands (CB). We use arrows as a guide to the eye for the convergence of the A peak.}
\end{figure}

\begin{figure}[ht!]
    \centering
    \includegraphics{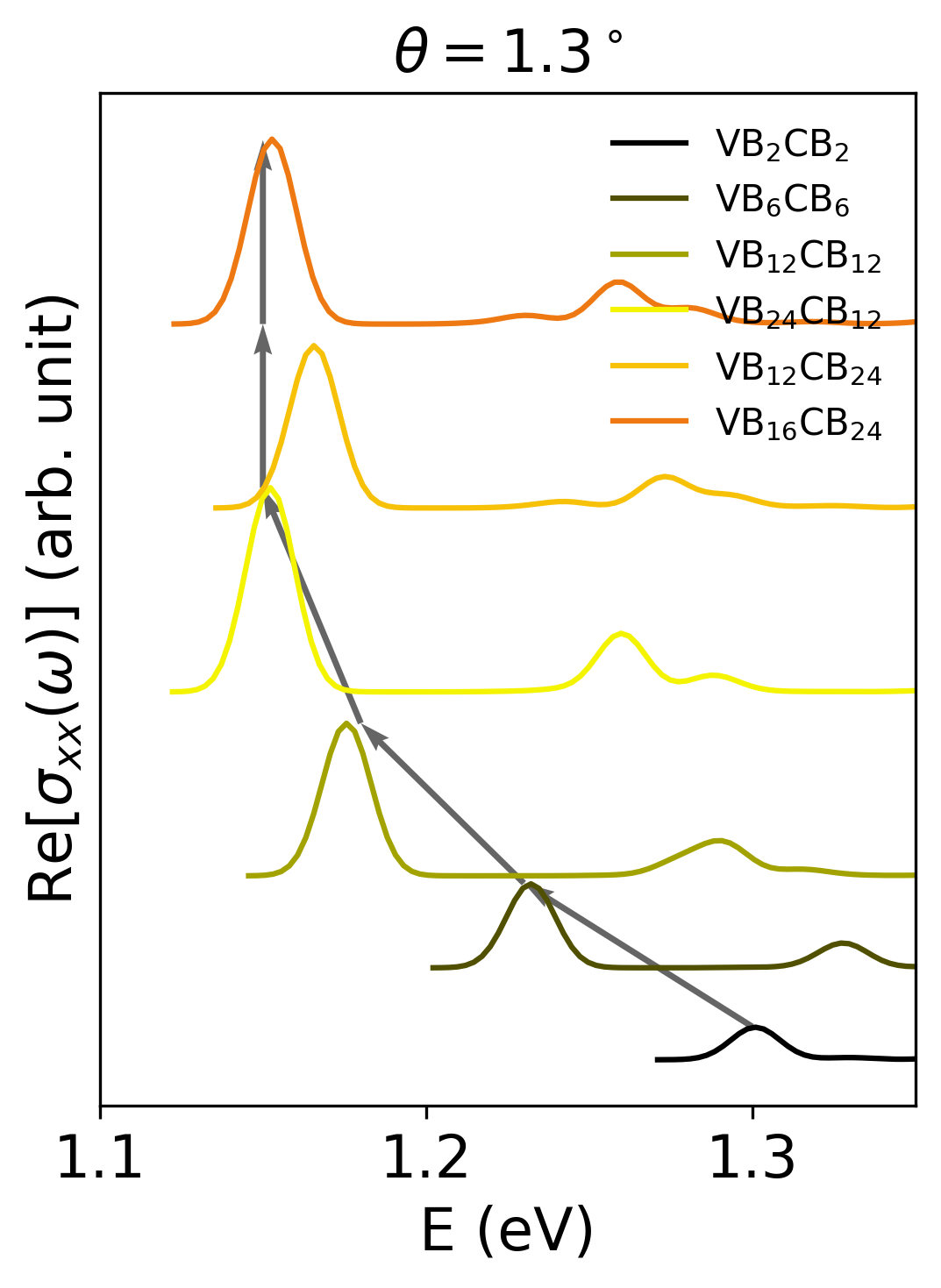}
    \caption{Impact of increasing the number of bands while construction of the BSE Hamiltonian in moir\'{e} systems. The figure shows the results of relaxed WSe$_{2}$ of a 1.3$^\circ$ twisted heterobilayer as representative of intralayer excitons of WSe$_{2}$. We do not include the GW correction and the spin-orbit coupling in the convergence study. We use a $3\times 3\times 1$ k-grid and vary the number of valence bands (VB) and conduction bands (CB). We use arrows as a guide to the eye for the convergence of the A peak.}
\end{figure}

\begin{figure}[ht!]
    \centering
    \includegraphics{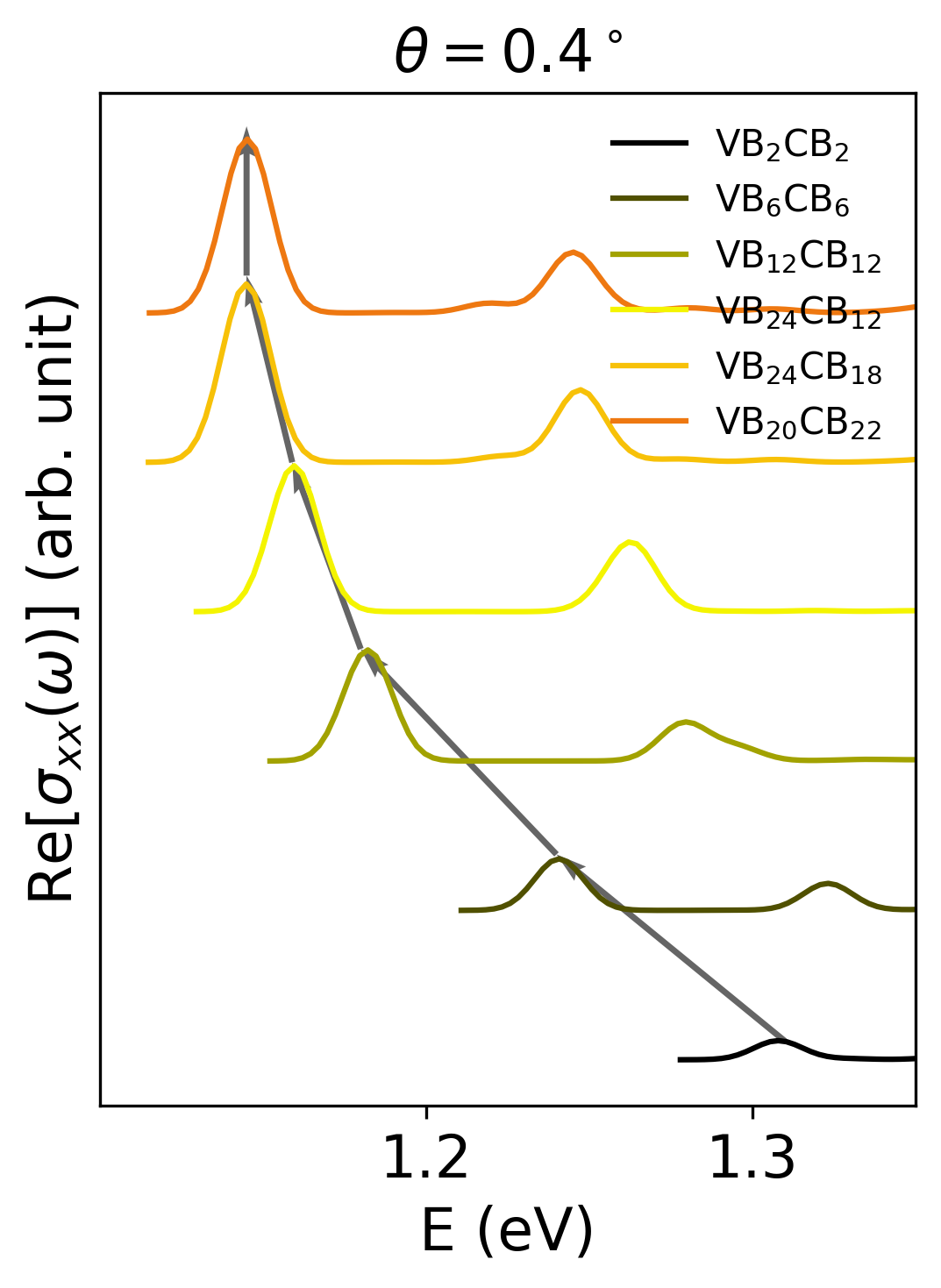}
    \caption{Impact of increasing the number of bands while construction of the BSE Hamiltonian in moir\'{e} systems. The figure shows the results of relaxed WSe$_{2}$ of a 0.4$^\circ$ twisted heterobilayer as representative of intralayer excitons of WSe$_{2}$. We do not include the GW correction and the spin-orbit coupling in the convergence study. We use a $3\times 3\times 1$ k-grid and vary the number of valence bands (VB) and conduction bands (CB). We use arrows as a guide to the eye for the convergence of the A peak.}
\end{figure}

\clearpage
\newpage

\subsection{IV. Impact of inclusion of the Exchange term during BSE Hamiltonian construction}
Previous studies have found that the inclusion of exchange interactions has a very small impact on the low-energy zero-momentum excitons~\cite{Fengchengexciton2015, Aghajanianoptical2023}. We compare the low-energy intralayer exciton optical conductivity of a monolayer WSe$_{2}$ and WSe$_{2}$ from a twisted WS$_{2}$/WSe$_{2}$ bilayer and find this to hold true in our work as well. 
\begin{figure}[ht!]
    \centering
    \includegraphics[scale=0.55]{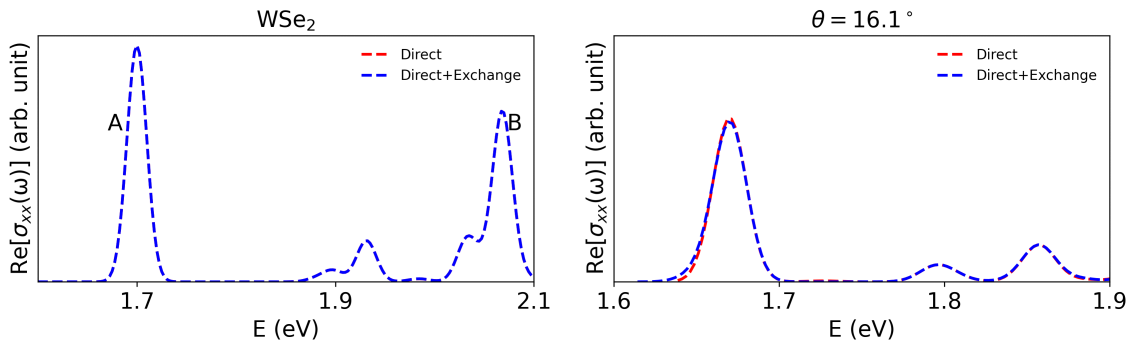}
    \caption{Comparison of the optical conductivity of a monolayer WSe$_{2}$ by including only the Direct term and both the Direct and Exchange terms. The monolayer WSe$_{2}$ BSE calculations were performed on a $39\times39\times1$ $k$-grid with 2 valence and 4 conduction bands. The twisted bilayer BSE calculations were performed on a $15\times15\times1$ $k$-grid with 4 valence and 4 conduction moir\'{e} bands.}
\end{figure}
\clearpage
\newpage

\subsection{V: Origin of slow convergence in moir\'{e} systems}
\begin{figure}[ht]
    \centering
    \includegraphics[scale=0.4]{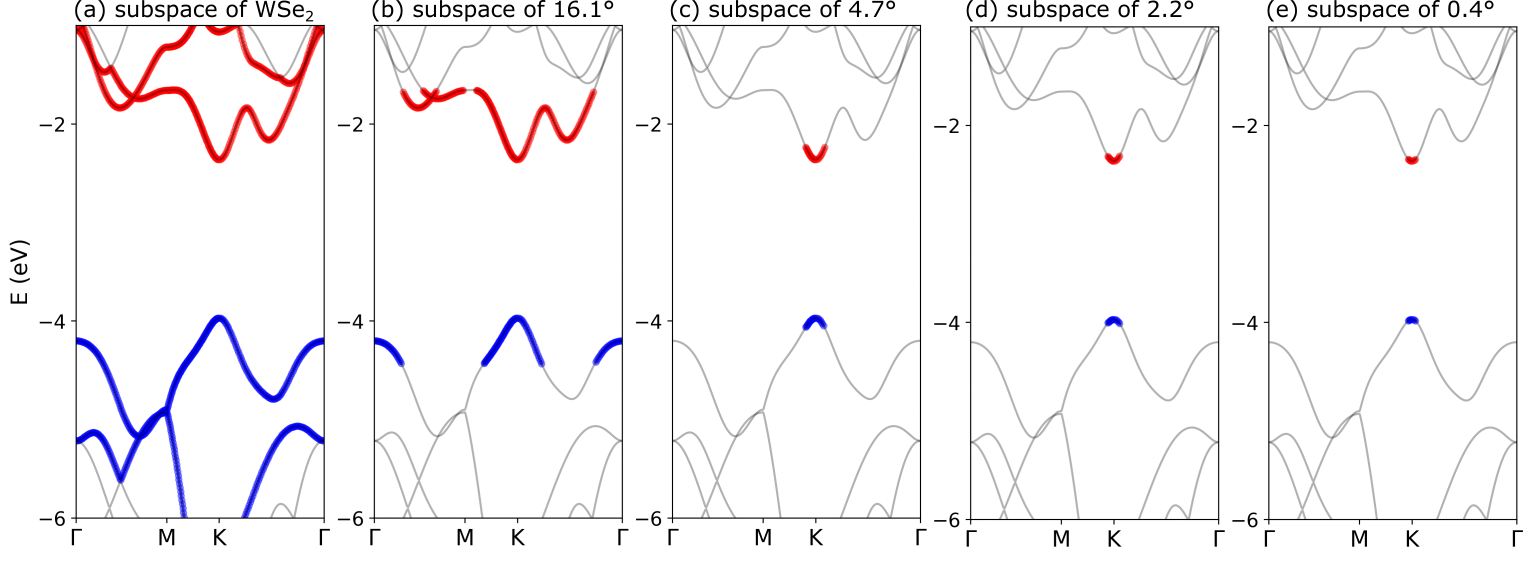}
      \includegraphics[scale=0.4]{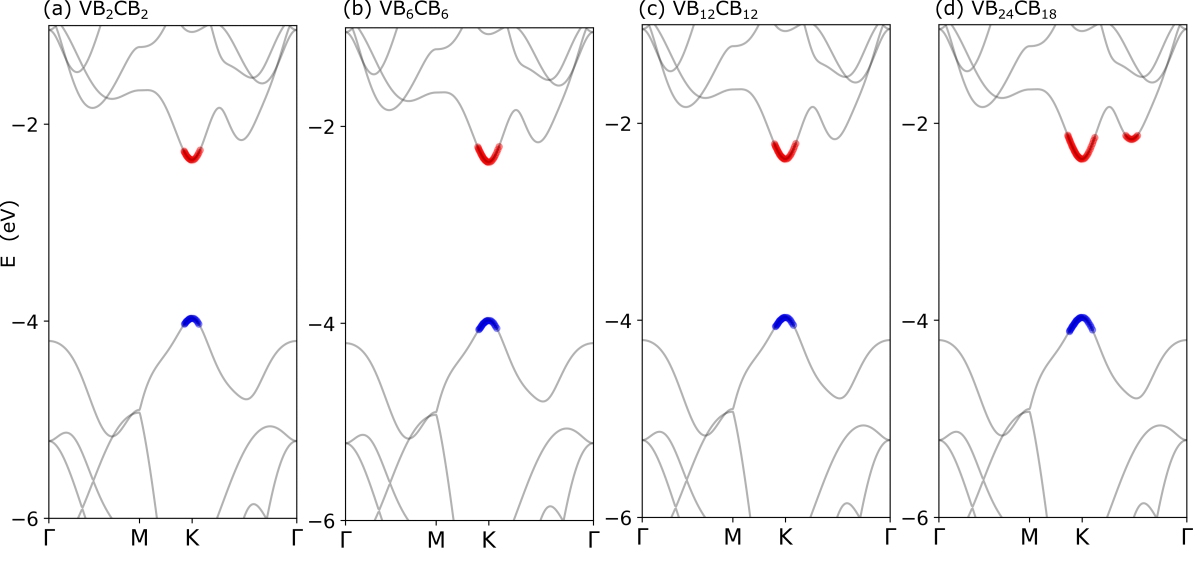}
    \caption{Top panels: Energies associated with 2 valence bands (marked by blue circles) and 2 conduction bands (marked by red circles) for different twist angles are shown. For small twist angles, a large number of valence and conduction bands are required to achieve convergence for low-lying excitons. This is a direct consequence of the small Brillouin zone in large scale moir\'{e} pattern.
    Bottom panels: Energies associated with incement of valence bands (marked by blue circles) and conduction bands (marked by red circles) for 0.4$^\circ$ moir\'{e} Brillouin zone.}
   \label{mbz}
\end{figure}

\subsection{VI: Exchange interaction: To screen or not to screen}
We have recalculated the optical conductivity of monolayer WSe$_{2}$ using the screened Keldysh interaction instead. As shown in Fig.~\ref{barevsscreened} below, it can be seen that the results are unaffected.  

\begin{figure}[ht]
    \centering
    \includegraphics[scale=0.75]{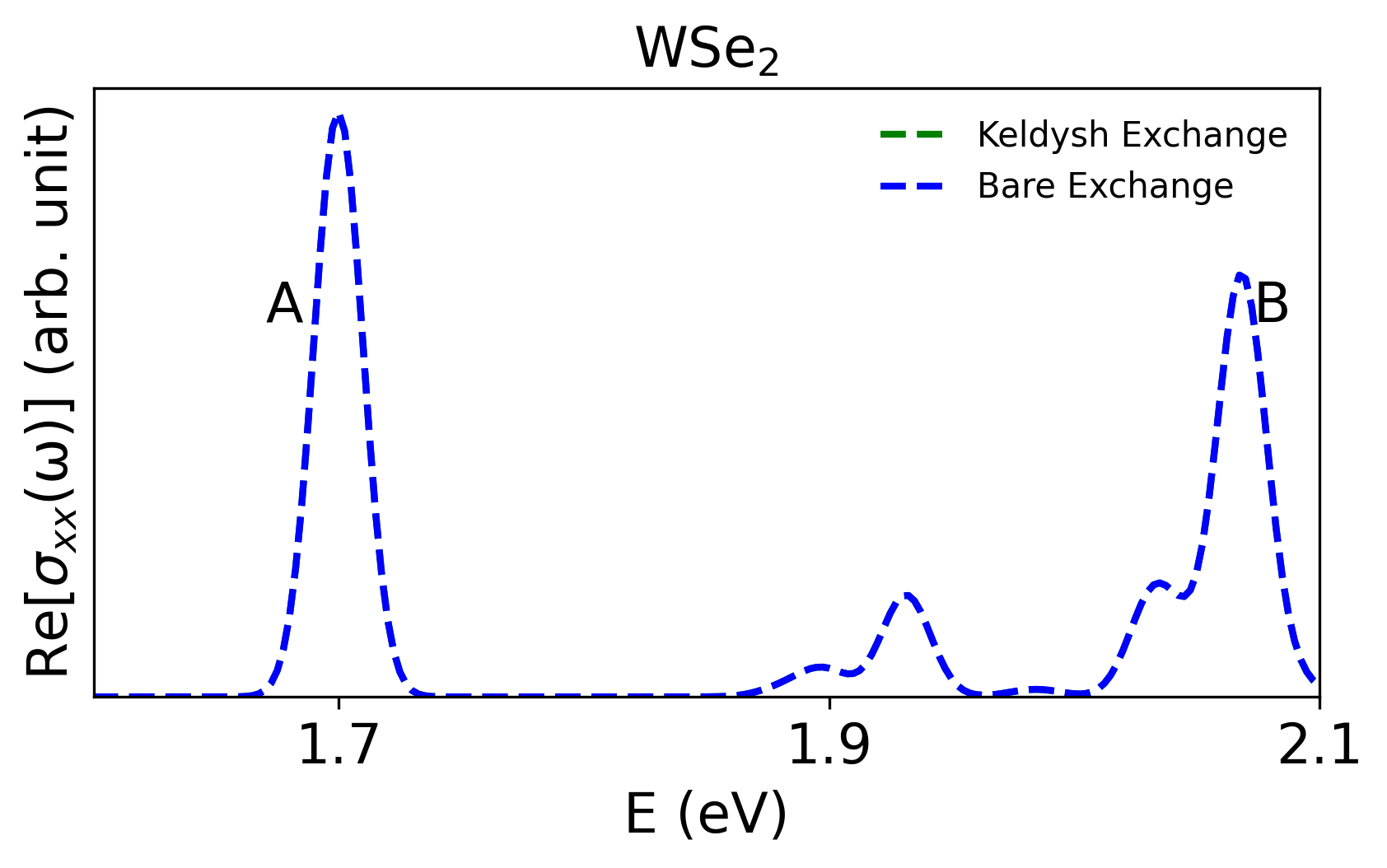}
    \caption{Optical conductivity of monolayer WSe$_{2}$ considering exchange interactions with both the bare Coulomb and Keldysh potentials. In both calculations, the Keldysh interaction is used for the direct contribution to the BSE kernel.}
   \label{barevsscreened}
\end{figure}

\clearpage
\newpage

\section{F: Electronic wavefunctions and contributions to excitons}
\begin{figure}[ht!]
    \centering
    \includegraphics[scale=0.3]{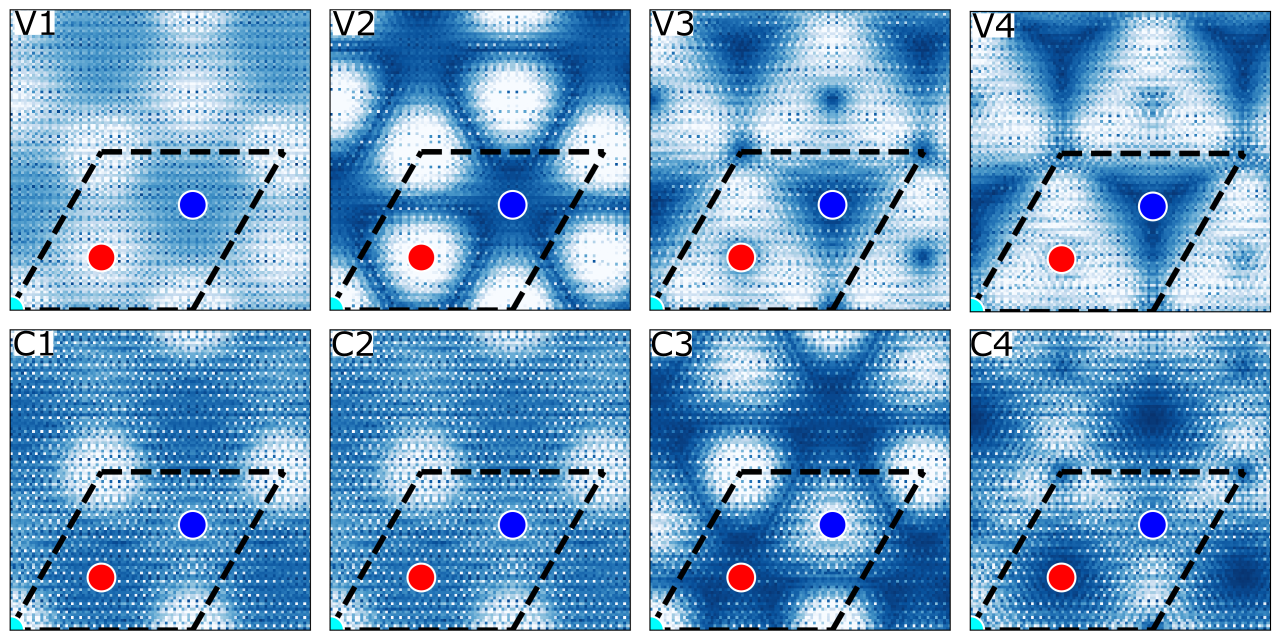}
    \caption{The squared absolute magnitudes of 3 valence (V1, V2, V3) and conduction (C1, C2, C3) band wave functions near the band gap at the $\Gamma$-point of the moiré unit cell for a $0^\circ$ twist angle with SZP as basis in SIESTA. The wave functions are averaged over the out-of-plane direction. We have used a moir\'{e} lattice constant of 7 nm. The different high-symmetry stacking regions in the moiré unit cell are indicated by symbols. V1 denotes the valence band maximum and C1 denotes the conduction band minimum.}
\end{figure}

\clearpage
\newpage 

\section{G: Optical conductivity of isolated WSe$_{2}$ and WS$_{2}$/WSe$_{2}$}
\begin{figure}[ht]
    \centering
    \includegraphics[scale=0.75]{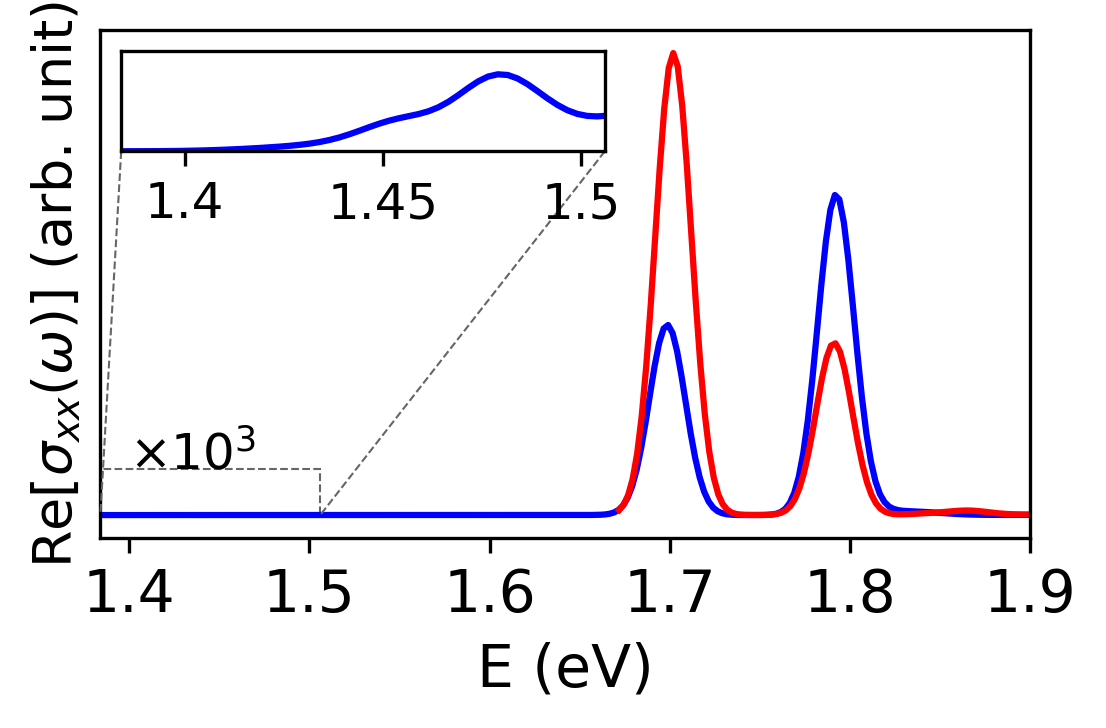}
    \caption{Optical conductivity of isolated WSe$_{2}$ (red solid line) and WS$_{2}$/WSe$_{2}$ heterobilayer (blue solid line) at a 5.7$^\circ$ twist angle. The BSE Hamiltonian for isolated WSe$_{2}$ was constructed with 12 valence and 12 conduction bands, while for the WS$_{2}$/WSe$_{2}$ heterobilayer, 12 valence and 28 conduction bands were used. While the GW shift in calculations for the isolated WSe$_{2}$ layer can be applied as a rigid shift, the same approach is not as straightforward for the heterobilayer, as the GW shifts for WS$_{2}$ and WSe$_{2}$ differ. In the above, we align only the 1s-like dominant intralayer excitons across different calculations.}
   \label{intrainter}
\end{figure}

\clearpage
\newpage

\section{H: Intralayer exciton wavefunctions of WSe$_{2}$}
\begin{figure}[ht]
    \centering
    \includegraphics[scale=0.85]{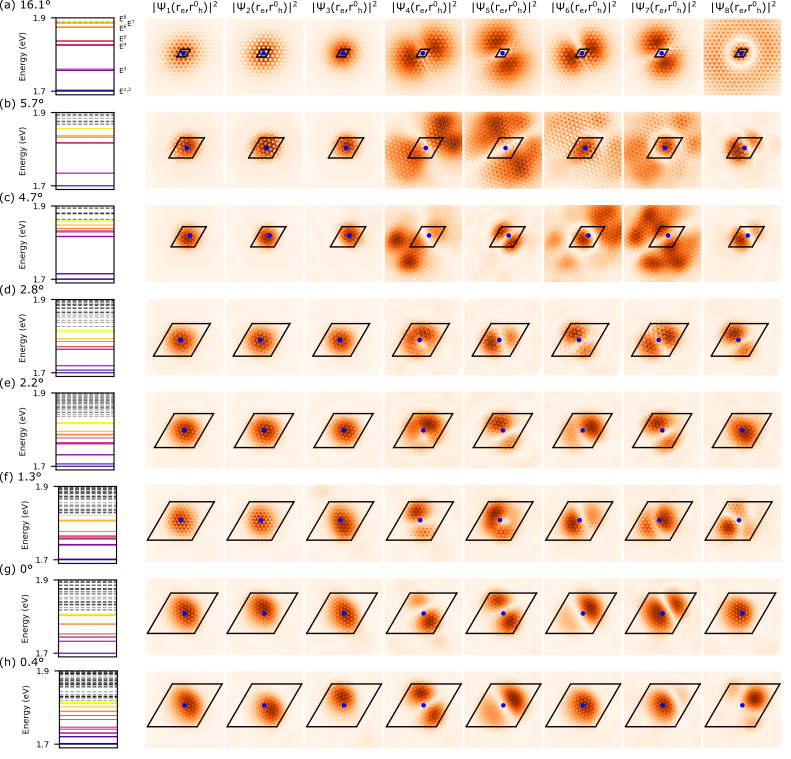}
    \caption{Energy-ordered intralayer exciton wavefunctions for each twist angle, with the hole located at the center (from left to right denotes an increase in energy). For exciton states with the same angular symmetry, we have summed over states with energies within approximately 5 meV.}
\end{figure}

\section{References}
\bibliography{exciton}
%\printbibliography
\end{document}